\documentclass[usegraphicx,useAMS]{mn2e}

\newcommand{\kms}{\mbox{${\rm km\,s}^{-1}$}}

\title[Li depletion in IC 4665]
  {Low-mass members of the young cluster IC~4665 and pre-main-sequence lithium depletion}
\author[R.~D. Jeffries et al.]
  {R.~D.~Jeffries$^{1}$, R.~J.~Jackson$^{1}$, David~J. James$^{2,3}$ and
  P.~A.~Cargile$^3$\\
$^{1}$  Astrophysics Group, Keele University, Keele, 
      Staffordshire ST5 5BG, UK\\
$^2$ Physics and Astronomy Department, University of Hawai'i, Hilo,
  Hawai'i, USA\\
$^3$ Department of Physics and Astronomy, Vanderbilt University,
  Nashville, Tenessee, USA
}
\date{Submitted 22 June 2009}

\pagerange{\pageref{firstpage}--\pageref{lastpage}} \pubyear{2004}

\def\LaTeX{L\kern-.36em\raise.3ex\hbox{a}\kern-.15em
    T\kern-.1667em\lower.7ex\hbox{E}\kern-.125emX}

\begin{document}

\label{firstpage}

\maketitle

\begin{abstract}
We have used fibre spectroscopy to establish cluster membership and
examine pre-main-sequence (PMS) lithium depletion for low-mass stars
(spectral types F to M) in the sparse young ($\simeq 30$\,Myr) cluster
IC~4665.  We present a filtered candidate list of 40 stars that should
contain 75 per cent of single cluster members with $11.5<V<18$ in the
central square degree of the cluster.  Whilst F- and G-type stars in
IC~4665 have depleted little or no lithium, the K- and early M-type
stars have depleted more Li than expected when compared with similar
stars in other clusters of known age. An empirical age estimate based
on Li-depletion among the late-type stars of IC~4665 would suggest it
is older than 100\,Myr. This disagrees entirely with ages determined
either from the nuclear turn-off, from isochronal matches to low-mass
stars or from the re-appearance of lithium previously found in much
lower mass stars (the ``lithium depletion boundary'').  We suggest that
other parameters besides age, perhaps composition or rotation, are very
influential in determining the degree of PMS Li-depletion in stars with
$M>0.5 M_{\odot}$. Further work is required to identify and assess the
effects of these additional parameters, particularly to probe
conditions at the interface between the sub-photospheric convection
zone and developing radiative core. Until then, PMS Li depletion in F-
to early M-type stars cannot be confidently used as a precise age
indicator in young clusters, kinematic groups or individual field
stars.
\end{abstract}


\section{Introduction}

The lithium depletion boundary (LDB) technique determines ages for
young clusters of stars by establishing the luminosity at which Li
remains unburned in the atmospheres of fully-convective, very low-mass
stars. In principle, LDB ages are both precise and accurate;
observational and theoretical uncertainties contribute to errors
of no more than about 10 per cent in the age estimate for
clusters in the range 10--200\,Myr (Jeffries \& Naylor 2001; Burke
et al. 2004) -- considerably better than other age
estimation methods.

LDB ages have now been estimated for several clusters (Stauffer et
al. 1998; Barrado y Navascu\'es, Stauffer \& Jayawardhana 2004;
   Jeffries \& Oliveira 2005), but the technique has limited applicability
because it entails quantifying the Li~{\sc i}~6708\AA\
feature at a resolving power $R\geq 3000$ in groups of faint, very low-mass
objects. However, the few LDB ages that are known could be used to calibrate other
age estimation methods that are feasible in more distant populations and
isolated field stars, but which rely on more uncertain stellar physics.

A prime example is the age-dependent degree of Li depletion seen in
more massive (and many magnitudes brighter) F--K stars. Li is burned in
pre-main-sequence (PMS) stars as their contracting cores reach
temperatures $\sim 2.5\times10^{6}$\,K. In fully-convective stars this
leads to rapid and almost total Li destruction on less than a
Kelvin-Helmholtz timescale. However, in the standard picture, PMS stars
with $M>0.5\,M_{\odot}$ form radiative cores that push outwards, so that
the base of the convection zone dips below the Li ignition temperature,
bringing an abrupt halt to surface Li depletion (e.g. Piau \&
Turck-Ch\`ieze 2002). The exact details and
in particular, isochrones of Li depletion versus mass or effective
temperature, are extremely sensitive to conditions at the base of the convection
zone and any additional mixing into the radiative core. As a
result, whilst Li depletion among F--K PMS stars has a long history of
use as an empirical age indicator among young clusters, kinematic
groups and even individual stars (e.g. Duncan \& Jones 1983; Stauffer,
Hartmann \& Barrado y Navascu\'es 1995;  Favata et al. 1998; Song et
al. 2000), the absolute ages determined in this
way have large theoretical uncertainties.

In this paper we discuss results from a programme to calibrate Li
depletion ages among F--K stars by observing such stars in clusters
with known LDB ages. We present photometry and spectroscopy of F--K
(and M) stars in IC~4665, an open cluster with a distance of $370\pm
50$\,pc, a reddening $E(B-V)=0.18\pm 0.05$\,mag (Hogg \& Kron 1955;
Crawford \& Barnes 1972 -- corresponding to $E(V-I)=0.23\pm 0.06$ and
$A_{V}=0.59\pm 0.16$, for the intrinsic colours of a low-mass PMS star
[Bessell et al. 1998]) and an unambiguous detection of the LDB which
yields an age of $28\pm 5$\,Myr (Manzi et al. 2008, hereafter M08),
where the modest age uncertainty is dominated by the distance
uncertainty.

\section{Observations and Analysis}

\begin{figure}
\centering
\includegraphics[width=80mm]{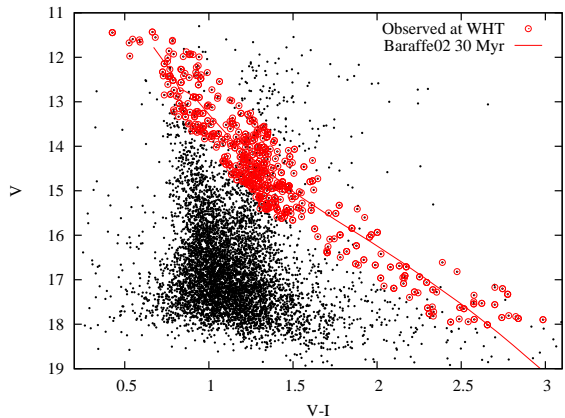}
\caption{A $V$ vs $V-I$ color magnitude diagram from our 
   photometric survey of IC~4665. Large symbols indicate targets
   (selected solely from this diagram) that were spectroscopically
   observed at the WHT. A 30\,Myr isochrone from the Baraffe et
   al. (2002) models is shown (using a colour/effective temperature
   relationship that is tuned using data from the Pleiades -- see Jeffries
   \& Oliveira [2005] for details), adjusted for a distance of 370\,pc,
   $A_V=0.59$\,mag and $E(V-I)=0.23$\,mag.}
\label{vvi}
\end{figure}

Spectroscopic observations of candidate low-mass stars in
IC~4665, with $11.5<V<18$ and $0.4<V-I<3$, were conducted on the nights beginning 8 June
2008 and 9 June 2008, using the AF2/WYFFOS multi-fibre spectrograph at a
Nasmyth focus of the 4.2-m William Herschel Telescope (WHT). An
echelle grating and an order-sorting filter were used to give a spectral coverage
of about 420\AA\ centred at 6600\AA\ and a resolving power of 11\,000.

Targets were selected from a $\simeq 1$ square degree photometric
survey centred at RA$=266.55^{\circ}$, Dec=$+5.7^{\circ}$ (J2000),
performed on 15--17 September 2005 using the 1.0-m telescope at the
Cerro Tololo Interamerican Observatory. The details of this survey will
appear elsewhere (Cargile et al. in prep.), but in brief consisted of
nine $19.3\times19.3$~arcmin$^2$ fields observed for 100, 50 and 25
seconds through $B,V,I$ filters. The observations were reduced with
{\sc daophot} (Stetson 1987) and placed onto the Johnson-Cousins system
using standards from Landolt (1992). The $V,V-I$ colour-magnitude
diagram (CMD) for the survey is shown in Fig.~1, along with targets
observed at the WHT. Targets were selected from this diagram assuming
the LDB age of 28\,Myr, but the criteria were broad enough to encompass
all likely cluster members and all previously identified low-mass
members in this magnitude range 
(e.g. Prosser \& Giampapa 1994 -- hereafter PG94; Mart\'in \&
Montes 1997 -- hereafter MM97). A total of 452
targets were observed in 8 fibre configurations, amounting to 87 per cent of possible
candidates with $11.5<V<18$. Four ``bright'' configurations were
observed for $3\times 600$\,s and 4 ``faint'' configurations were
observed for $3\times 1800$\,s (one had only two good exposures). The
spectra had signal-to-noise ratios (SNR) ranging from 4--110 per 0.22\AA\
pixel. 124 targets were observed twice, 37 targets were observed three
times and 6 targets were observed in four configurations.

Spectra were extracted from the CCD images using purpose built software
(written by RJJ) which used an optimal extraction procedure
to maximize the signal to noise ratio. Science images were corrected
for bias by subtracting a median of multiple bias frames and correcting
for any time-dependent offset using overscan regions. Bad pixels
were flagged using a mask generated from dark frames.  Tungsten lamp
images recorded before each science run were used to trace the path of
the spectrum on the CCD image. The median of these images was used to
define flat-field spectra for the optimal extraction
process. Copper-neon lamp spectra, recorded before and after science
exposures, provided the wavelength calibration.

The method used for optimal extraction is described by Horne
(1986). This applies non-uniform weight to pixels in the extracted sum,
minimizing statistical noise whist preserving photometric accuracy. In
our case the weighting profiles for individual spectra were derived
from a boxcar average along the wavelength axis of the flat-field
images. A similar boxcar averaging process using data from regions
between the spectra was used to model and subtract any scattered
light in each exposure. With this procedure the extracted science spectra were
normalized to the flat-field spectra, compensating for variations in
gain between pixels. To reduce the effect of cosmic rays the measured
signal profile was compared with the product of the extracted spectrum
and the weighting profile. Pixels showing greater than 5-sigma
difference were masked and the extraction process repeated.  Profiles
with more than 50 per cent of the pixels masked were masked in the final
extracted spectrum.

Arc spectra, extracted using the same procedure, were used to identify
the position of seven well separated arc lines and a cubic polynomial
fit was used to define the calibration of individual spectra. Extracted
spectra and their variances were interpolated across masked points
and recast onto a common wavelength base comprising 10\,000
logarithmically spaced steps between 6350\AA, and 6740\AA. A median sky
spectrum, determined from ten or more fibers pointed at sky positions
in each configurations was subtracted from the target spectra within
that configuration.  Spectra of the ``blank'' night sky were taken to
measure the relative efficiency of individual fibers in each
configuration. The results of repeated exposures were averaged to
produce the median spectrum of each target from each configuration.

Spectra of late-type stars with known radial velocities (RVs) 
(from Nidever et al 2002) were taken on each night.
Heliocentric RVs were measured by cross-correlation using the {\sc
iraf} {\sc fxcor} task and HD 171067 (spectral type G8V) as the
principal template. The cross-correlation used the wavelength regions
6380--6550\AA\ and 6580--6725\AA. Projected equatorial velocities ($v \sin i$) were
estimated from the width of the cross-correlation function. Translation
from the cross-correlation function width to a $v \sin i$ was
calibrated by artificially broadening standard star spectra. Variation
in the intrinsic resolution of different fibres was accounted for using the
cross-correlation function between HD~171067 and the twilight sky
spectrum recorded through each fibre.  At this resolution the data were
capable of resolving rotational broadening corresponding to $v \sin i>
20$\,km\,s$^{-1}$. {\sc fxcor} provides error estimates for the RVs,
which concurred with estimates from targets observed in multiple
configurations. The average RV uncertainty was 0.8\,km\,s$^{-1}$, but
varied from 0.3\,km\,s$^{-1}$ to as large as 8\,km\,s$^{-1}$ in the
most rapid rotators.  RVs were obtained for 449/452 targets, the
remaining 3 targets had low SNR and no discernible cross-correlation peak.  $v \sin
i$ errors were estimated by numerical simulation and were of order 10
per cent, where rotational broadening was resolved.

The equivalent widths of the Li\,{\sc i}~6707.8\AA\ feature (EW[Li])
were measured with respect to a pseudo-continuum, using a Gaussian fit
to the line itself where present. Uncertainties were estimated using
the empirical SNR of the continuum fit and the
formula $\sigma {\rm EW} = 1.5\sqrt{pw}/{\rm SNR}$, where $p$ is the pixel
size and $w$ the FWHM of the line in wavelength units (Cayrel 1988).  Where no
significant line was seen we estimated very conservative 3-sigma upper limits using this
formula and an assumed $w$ (including any rotational broadening). At
our resolution, the measured EW[Li] includes a blended contribution from
a weak ($\simeq 10$ to 20\,m\AA) Fe\,{\sc i} line at 6707.44\AA.

\begin{figure*}
\centering
\begin{minipage}[t]{0.45\textwidth}
\includegraphics[width=80mm]{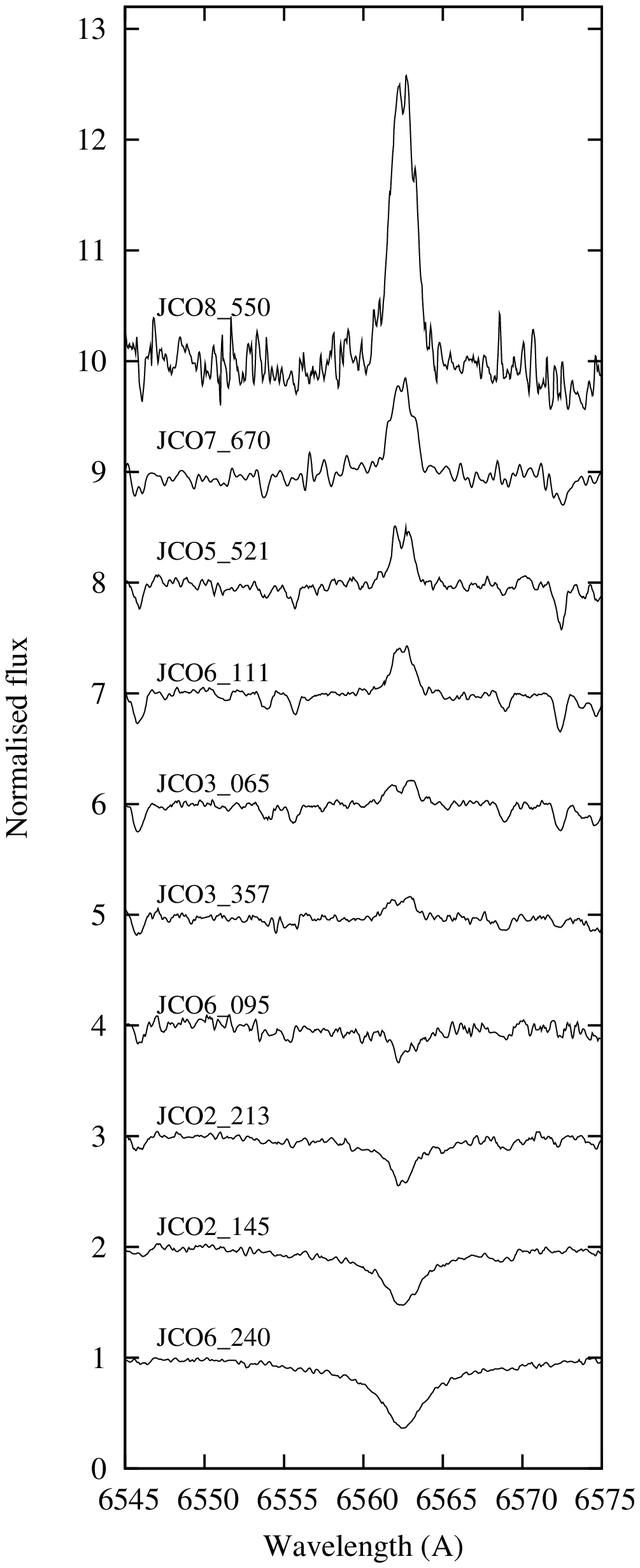}
\end{minipage}
\begin{minipage}[t]{0.45\textwidth}
\includegraphics[width=80mm]{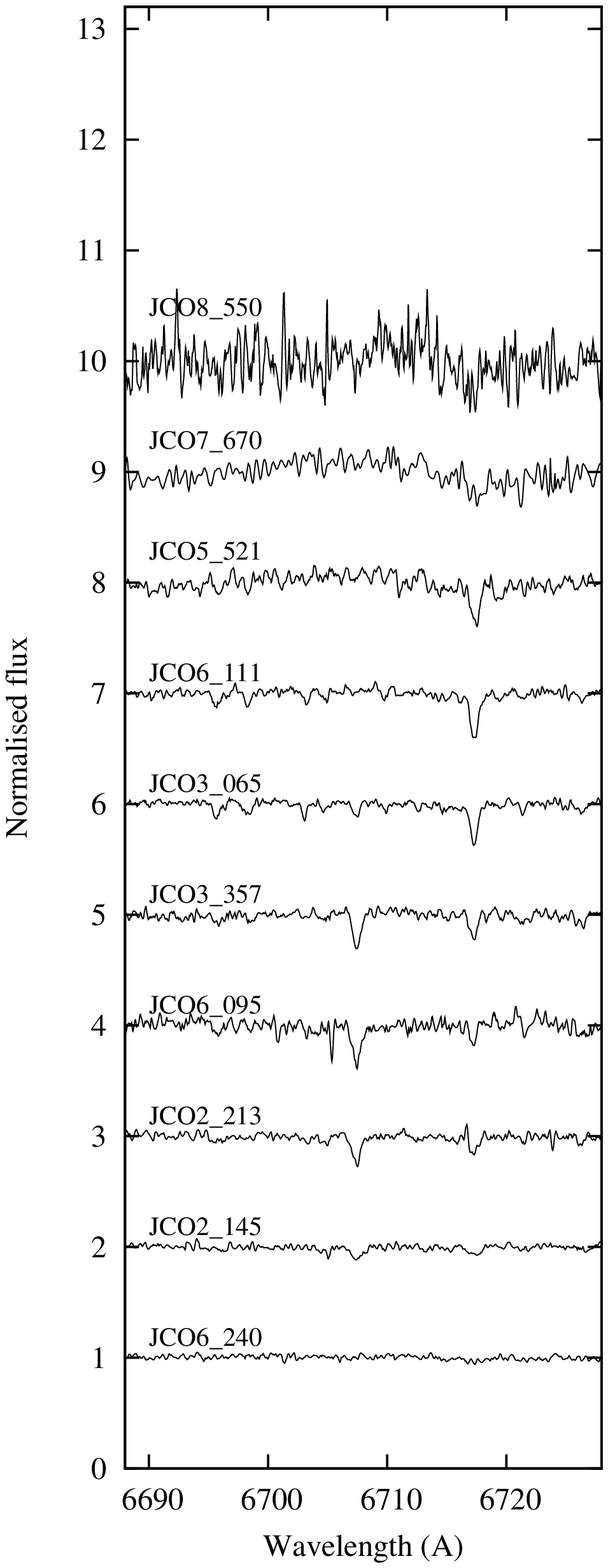}
\end{minipage}
\caption{Examples of spectra in the vicinities of the
  H$\alpha$ and Li\,{\sc i} 6708\AA\ lines. The spectra have been
  normalised to unity and offset for clarity. These examples show the
  full range of signal-to-noise ratios and spectral types for some
  candidate members of IC~4665 (see Table~2).
}
\label{spectra}
\end{figure*}

A summary of the results is included as Table~1 (available fully in
electronic form only). This identifies each target, gives its J2000
celestial coordinates (taken from the 2MASS catalogue -- Cutri et
al. 2003), its $V$ and $V-I$ magnitudes, the heliocentric RV
measurements, a weighted mean RV, a RV variability flag, a flag
identifying objects that are candidate SB2 binary systems based on the
appearance of the cross-correlation functions, the $v\sin i$ (or upper
limit), a SNR for the (summed) spectrum and a EW[Li]
(or upper limit) found from the (summed) spectrum of the
object. Table~1 also lists alternative star designations from Prosser
(1993) and M08.
Figure~\ref{spectra} shows a sample of our spectra
around the H$\alpha$ and Li~{\sc i}~6708\AA\ features.

\begin{table*}
\caption{The spectroscopic measurements of 452 photometric candidates
  in IC~4665. Only a few rows are shown here to illustrate the
  content, the full table is available in electronic form. Column 1 is
  an identifier from the photometric catalogue (Cargile et al. in
  prep), columns 2 and 3 give the RA and Dec (J2000, from the 2MASS
  catalogue), columns 4 and 5 give the $V$ and $V-I$ photometry, column
  6 lists how many observations were made of the target then columns
  7--9, 10--12, 13--15 and 16--18 give up to 4 possible measurements of
  the heliocentric RV, its error and the HJD of the
  observation. Columns 19 and 20 give the weighted mean RV and its
  uncertainty, column 21 gives the mean signal-to-noise ratio of the
  averaged spectrum, column 22 and 23 give the measured
  EW[Li] and its uncertainty. An uncertainty of -9999 indicates
  that the quoted value is a 3-sigma upper limit. Columns 24 and 25
  list the projected equatorial velocity and its uncertainty. An
  uncertainty of -9999 indicates that the $v \sin i$ is an upper
  limit. Column 26 indicates whether the object has a variable RV ($=1$
  if variable at
  $>99.9$ per cent confidence) and column 27 indicates whether the
  cross-correlation function with the template standard star shows
  evidence for more than one star ($=1$ for evidence of
  binarity). Columns 28 and 29 list alternative names for the targets
  from Prosser (1993) and Manzi et al. (2008).}
\begin{tabular}{cc@{\hspace*{1mm}}cccc@{\hspace*{1mm}}c@{\hspace
*{1mm}}c@{\hspace*{1mm}}cc@{\hspace*{1mm}}c@{\hspace*{1mm}}cc@{\hspace*{1mm}}c@{\hspace*{1mm}}c}
\hline
 (1)  & (2) & (3) & (4) & (5) &(6) &(7)&(8)&(9)&(10)& (11) & (12) &
(13) & (14) & (15) \\
Identifier & RA & Dec & $V$ & $V-I$ & $N_{\rm obs}$ &
\multicolumn{3}{c}{Measurement 1} &\multicolumn{3}{c}{Measurement 2}&
\multicolumn{3}{c}{Measurement 3}\\ 
      &    \multicolumn{2}{c}{(J2000)}& & &  &
 HJD& RV
  & $\sigma$RV & HJD& RV & $\sigma$RV &HJD& RV & $\sigma$RV \\ 
      &     &     &     &     & & $-2454600$ &\multicolumn{2}{c}{(km/s)}  &$-2454600$ &\multicolumn{2}{c}{(km/s)}  &$-2454600$ &\multicolumn{2}{c}{(km/s)}  \\
\hline
 JCO1\_025 & 266.89118 &  6.16281& 15.303 & 1.313& 1&    27.557&   39.8&
 0.5&  &  & &&&\\        
 JCO1\_237 & 266.88043 &  6.03657& 14.337 & 1.299& 2&    26.469&   12.0&   
 0.7&   27.424&    10.7 &    0.6 & & &\\
JCO3\_265 & 266.39267 &  5.97103& 14.800&  1.438& 4&    26.535&  -47.1&    0.5&    27.478&   -47.2&     0.8&    27.557&   -46.7&     0.5 \\
\multicolumn{15}{c}{...}\\
\hline
\end{tabular}
\begin{tabular}{cc@{\hspace*{1mm}}c@{\hspace*{1mm}}cc@{\hspace*{1mm}}ccc@{\hspace*{1mm}}cc@{\hspace*{1mm}}ccccc}\\
\hline
(1) & (16)& (17)& (18)& (19)& (20)& (21)& (22)& (23)& (24)& (25)&
(26)& (27) & (28) & (29)\\
Identifier & \multicolumn{3}{c}{Measurement 4} & Mean RV & $\sigma$RV&
SNR & EW[Li] & $\sigma$EW[Li] & $v\sin i$ & $\sigma(v \sin i)$ & Var &
CCF & P93 & M08 \\
    &  HJD& RV & $\sigma$RV & & & & & & & & & & & \\
    & $-2454600$ &\multicolumn{2}{c}{(km/s)} & \multicolumn{2}{c}{(km/s)}
& & \multicolumn{2}{c}{(m\AA)}&\multicolumn{2}{c}{(km/s)} & & & & \\ 
\hline
 JCO1\_025 & & & & 39.8 & 0.5 & 18 &  108& -9999&   20.0&   -9999&  0& 0&&\\
 JCO1\_237 & & & & 11.3 & 0.4 & 29 &   68& -9999&   20.0&   -9999&  0& 0&&\\
 JCO3\_265 & 27.628&   -46.3&     0.7&    -46.8&  0.3&  46&  42& -9999&   20.0&   -9999&  0& 0&&\\
\multicolumn{15}{c}{...}\\
\hline
\end{tabular}
\label{tab1}
\end{table*}

\section{Cluster Membership}

\label{membership}

\begin{figure}
\centering
\includegraphics[width=80mm]{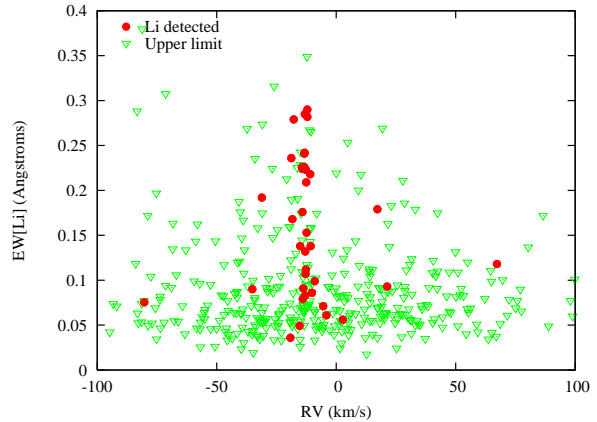}
\caption{The measured equivalent width of the Li\,{\sc i}~6708\AA\
  feature aginst heliocentric radial velocity. Solid points are
  detections, whereas triangles represent 3-sigma upper limits. There is
  a clear concentration of Li-detections at an RV $\simeq -13$\,km\,s$^{-1}$.}
\label{lirv}
\end{figure}

\begin{figure}
\centering
\includegraphics[width=80mm]{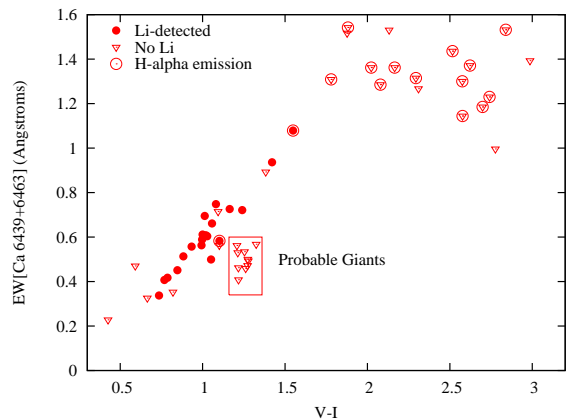}
\caption{The summed equivalent width of the Ca\,{\sc i} lines at
   6349\AA\ and 6463\AA\ as a function of colour. Candidates with a
   detected Li\,{\sc i} feature or H$\alpha$ in emission are
   indicated. The candidates in the box have weak Ca\,{\sc i} and we
   class them as probable giants and hence non-members.}
\label{cavi}
\end{figure}

\begin{figure}
\centering
\includegraphics[width=80mm]{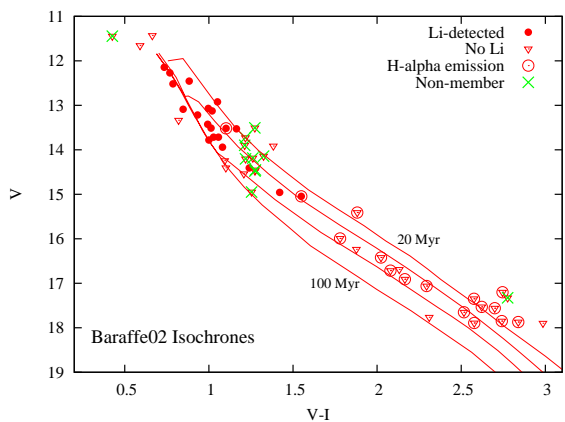}
\caption{A $V$ vs $V-I$ color magnitude diagram 
   for the 56 stars in IC~4665 with 
   RVs compatible with cluster membership. Stars which have Li in
   absorption or H$\alpha$ emission are indicated, as are 12 stars
   judged to be non-members on the basis of their giant-like
   gravities or discrepant proper-motions (see text). Baraffe et al. (2002)
   isochrones at 20, 30, 50 and 100\,Myr are also shown (these use a
   Pleiades-tuned colour/effective temperature relationship -- see
   Jeffries \& Oliveira [2005]).
}
\label{vvi2}
\end{figure}

\begin{table*}
\caption{Properties of candidate RV members of IC~4665. Columns 1--6
  identify the star and give its photometry, EW[Li], RV and $v
  \sin i$. An uncertainty of -9999 indicates an upper limit. Column 7
  gives the EW of the H$\alpha$ feature (positive $=$ emission). Columns 8
  and 9 give the EWs of the gravity-sensitive Ca\,{\sc i} lines and
  columns 10 and 11 give the proper motion from the {\sc nomad}
  catalogue. Column 12 gives our final assessment of membership (see section~\ref{membership}).
  }
\begin{tabular}{cccr@{\hspace*{1mm}}rr@{\hspace*{1mm}}rr@{\hspace*{1mm}}r@{\hspace*{1mm}}cccr@{\hspace*{1mm}}rr@{\hspace*{1mm}}r@{\hspace*{1mm}}c}
\hline
 (1)  & (2) & (3) & \multicolumn{2}{c}{(4)} & \multicolumn{2}{c}{(5)} &\multicolumn{2}{c}{(6)} &(7)&(8)&(9)& \multicolumn{2}{c}{(10)}&  \multicolumn{2}{c}{(11)} & (12) \\
Identifier & $V$ & $V-I$ & \multicolumn{2}{c}{EW[Li]} 
& \multicolumn{2}{c}{RV}  & \multicolumn{2}{c}{$v\sin i$} & EW[H$\alpha$] &
\multicolumn{2}{c}{EW[Ca\,{\sc i}] (\AA)} &  \multicolumn{2}{c}{PM(RA)}  & \multicolumn{2}{c}{PM(Dec)}
  & Member \\ 
           &     &       &  \multicolumn{2}{c}{(m\AA)} &
\multicolumn{2}{c}{(km\,s$^{-1}$)} & \multicolumn{2}{c}{(km\,s$^{-1}$)}
& (\AA) &  6439 & 6463 & \multicolumn{4}{c}{(milli-arcsec year$^{-1}$)}\\
\hline
 JCO1\_427 & 13.07& 1.00&  218 $\pm$&     24&  -11.4 $\pm$&   2.5&  57.0 $\pm$&      2.2&  -0.75&   0.27&   0.32&    0.2 $\pm$&  5.9& -12.7 $\pm$&  5.9&   Y\\
 JCO1\_530 & 12.52& 0.79&   79 $\pm$&      6&  -14.5 $\pm$&   0.3&  20.0 $\pm$&    -9999&  -2.33&   0.18&   0.24&  -10.4 $\pm$&  3.0&  -1.5 $\pm$&  2.8&   Y\\
 JCO2\_026 & 14.95& 1.25&   66 $\pm$&  -9999&  -14.2 $\pm$&   0.5&  20.0 $\pm$&    -9999&  -1.11&   0.21&   0.33&    1.4 $\pm$&  5.9&   1.2 $\pm$&  6.1&   N\\
 JCO2\_145 & 12.27& 0.77&  153 $\pm$&     18&  -13.0 $\pm$&   1.7&  42.3 $\pm$&      2.2&  -2.95&   0.16&   0.24&   -5.3 $\pm$&  2.9& -11.2 $\pm$&  2.8&   Y\\
 JCO2\_213 & 13.22& 0.93&  226 $\pm$&     24&  -14.9 $\pm$&   0.7&  20.0 $\pm$&    -9999&  -1.19&   0.26&   0.29&   -2.5 $\pm$&  5.9&  -8.9 $\pm$&  5.9&   Y\\
 JCO2\_220 & 16.42& 2.02&  144 $\pm$&  -9999&  -14.9 $\pm$&   1.0&  21.0 $\pm$&    -9999&   1.16&   0.72&   0.64&             &     &            &     &   Y\\
 JCO2\_373 & 12.46& 0.88&  176 $\pm$&     15&  -14.7 $\pm$&   0.5&  20.0 $\pm$&    -9999&  -1.13&   0.24&   0.27&   -3.2 $\pm$&  1.5&  -6.3 $\pm$&  3.5&   Y\\
 JCO2\_637 & 17.20& 2.75&  435 $\pm$&  -9999&  -14.2 $\pm$&   4.0&  75.8 $\pm$&      7.6&   3.45&   0.88&   0.80&   -8.0 $\pm$&  1.0&   2.0 $\pm$&  5.0&   Y\\
 JCO2\_736 & 17.90& 2.99&  264 $\pm$&  -9999&  -11.2 $\pm$&   3.0&  20.0 $\pm$&    -9999&  -0.01&   0.80&   0.60&   14.0 $\pm$&  5.0&  -6.0 $\pm$&  4.0&   ?\\
 JCO3\_065 & 15.04& 1.55&   91 $\pm$&     22&  -14.3 $\pm$&   0.6&  20.0 $\pm$&    -9999&   0.32&   0.55&   0.53&    2.0 $\pm$&  5.9&  -9.5 $\pm$&  5.9&   Y\\
 JCO3\_285 & 13.34& 0.82&   72 $\pm$&  -9999&  -14.3 $\pm$&   0.9&  20.0 $\pm$&    -9999&  -2.12&   0.17&   0.18&   -0.8 $\pm$&  3.0&  -7.7 $\pm$&  2.9&   Y\\
 JCO3\_357 & 13.52& 1.10&  290 $\pm$&     27&  -12.6 $\pm$&   0.7&  20.0 $\pm$&    -9999&   0.34&   0.27&   0.32&   -2.2 $\pm$&  5.9&  -3.6 $\pm$&  5.9&   Y\\
 JCO3\_395 & 13.72& 1.06&  282 $\pm$&     19&  -12.7 $\pm$&   0.4&  20.0 $\pm$&    -9999&  -0.28&   0.33&   0.33&    1.3 $\pm$&  5.9&  -8.5 $\pm$&  5.9&   Y\\
 JCO3\_396 & 13.43& 0.99&  209 $\pm$&     18&  -13.1 $\pm$&   0.4&  20.0 $\pm$&    -9999&  -1.27&   0.28&   0.29&    4.0 $\pm$&  5.9&  -3.7 $\pm$&  5.9&   Y\\
 JCO3\_724 & 17.77& 2.31&  348 $\pm$&  -9999&  -12.8 $\pm$&   2.3&  23.0 $\pm$&    -9999&  -0.23&   0.64&   0.63&   16.0 $\pm$&  6.0&   0.0 $\pm$&  3.0&   ?\\
 JCO3\_770 & 17.56& 2.70&  225 $\pm$&  -9999&  -16.8 $\pm$&   2.9&  22.2 $\pm$&      4.9&   3.58&   0.65&   0.53&    2.0 $\pm$&  2.0&  -8.0 $\pm$&  0.0&   Y\\
 JCO4\_053 & 13.53& 1.16&  226 $\pm$&     19&  -13.9 $\pm$&   0.9&  32.5 $\pm$&      1.7&  -0.10&   0.38&   0.35&   -0.2 $\pm$&  5.9&  -9.5 $\pm$&  5.9&   Y\\
 JCO4\_149 & 14.19& 1.26&   66 $\pm$&  -9999&  -13.8 $\pm$&   0.4&  20.0 $\pm$&    -9999&  -1.03&   0.18&   0.28&   -3.8 $\pm$&  5.9& -10.9 $\pm$&  5.9&   N\\
 JCO4\_226 & 13.78& 1.00&  224 $\pm$&     19&  -14.7 $\pm$&   0.7&  20.0 $\pm$&    -9999&  -0.96&   0.30&   0.31&   -4.4 $\pm$&  5.9&  -6.1 $\pm$&  5.9&   Y\\
 JCO4\_295 & 11.45& 0.43&   42 $\pm$&  -9999&   -6.4 $\pm$&   7.6&  82.5 $\pm$&      6.1&  -8.20&   0.12&   0.11&  -13.0 $\pm$&  2.5&   1.6 $\pm$&  2.4&   N\\
 JCO4\_337 & 13.91& 1.38&   69 $\pm$&  -9999&  -13.9 $\pm$&   0.5&  20.0 $\pm$&    -9999&  -0.89&   0.43&   0.46&    8.6 $\pm$&  5.9&   3.9 $\pm$&  5.9&   Y\\
 JCO4\_437 & 16.72& 2.08&  142 $\pm$&  -9999&  -11.5 $\pm$&   1.0&  20.0 $\pm$&    -9999&   0.35&   0.60&   0.68&             &     &            &     &   Y\\
 JCO4\_459 & 17.89& 2.58&  190 $\pm$&  -9999&  -16.7 $\pm$&   2.4&  20.0 $\pm$&    -9999&   3.84&   0.53&   0.61&             &     &            &     &   Y\\
 JCO4\_558 & 17.33& 2.78&  174 $\pm$&  -9999&   -8.7 $\pm$&   4.5&  25.0 $\pm$&    -9999&  -0.20&   0.50&   0.50&  -26.0 $\pm$&  4.0& -34.0 $\pm$&  2.0&   N\\
 JCO4\_591 & 16.24& 1.88&   69 $\pm$&  -9999&  -12.4 $\pm$&   0.5&  20.0 $\pm$&    -9999&  -0.60&   0.77&   0.75&    3.1 $\pm$&  6.2&  -7.2 $\pm$&  6.3&   Y\\
 JCO5\_129 & 13.73& 1.22&   49 $\pm$&  -9999&  -13.9 $\pm$&   0.3&  20.0 $\pm$&    -9999&  -1.06&   0.19&   0.27&   -1.1 $\pm$&  5.9&   2.2 $\pm$&  5.9&   N\\
 JCO5\_179 & 14.41& 1.24&  242 $\pm$&     49&  -13.8 $\pm$&   0.8&  20.0 $\pm$&    -9999&  -0.04&   0.33&   0.39&   -0.1 $\pm$&  5.9&  -3.1 $\pm$&  5.9&   Y\\
 JCO5\_280 & 14.24& 1.09&   70 $\pm$&  -9999&  -14.4 $\pm$&   0.3&  20.0 $\pm$&    -9999&  -0.79&   0.30&   0.41&    0.0 $\pm$&  5.9&  -6.9 $\pm$&  5.9&   Y\\
 JCO5\_282 & 12.15& 0.73&  112 $\pm$&      9&  -13.2 $\pm$&   1.1&  34.8 $\pm$&      1.8&  -1.95&   0.15&   0.18&    1.8 $\pm$&  1.5&  -7.4 $\pm$&  1.5&   Y\\
 JCO5\_296 & 13.94& 1.08&  279 $\pm$&     64&  -18.4 $\pm$&   3.4&  72.0 $\pm$&      2.9&  -0.01&   0.35&   0.40&    8.0 $\pm$&  5.9&  -5.1 $\pm$&  5.9&   Y\\
 JCO5\_329 & 11.43& 0.66&   25 $\pm$&  -9999&  -15.6 $\pm$&   1.1&  41.9 $\pm$&      2.0&  -3.84&   0.17&   0.15&    2.8 $\pm$&  4.3&  -7.2 $\pm$&  2.7&   ?\\
 JCO5\_394 & 16.69& 2.13&  114 $\pm$&  -9999&  -12.3 $\pm$&   0.7&  20.0 $\pm$&    -9999&  -0.31&   0.79&   0.74&             &     &            &     &   ?\\
 JCO5\_472 & 16.91& 2.17&  102 $\pm$&  -9999&  -14.4 $\pm$&   0.6&  20.0 $\pm$&    -9999&   0.74&   0.71&   0.65&             &     &            &     &   Y\\
 JCO5\_515 & 17.85& 2.74&  228 $\pm$&  -9999&  -15.4 $\pm$&   2.3&  23.0 $\pm$&    -9999&   1.93&   0.60&   0.63&             &     &            &     &   Y\\
 JCO5\_521 & 17.06& 2.29&  108 $\pm$&  -9999&  -13.6 $\pm$&   1.0&  20.0 $\pm$&    -9999&   0.66&   0.70&   0.62&             &     &            &     &   Y\\
 JCO6\_018 & 13.90& 1.21&   70 $\pm$&  -9999&  -12.8 $\pm$&   0.6&  20.0 $\pm$&    -9999&  -1.04&   0.24&   0.29&   -2.9 $\pm$&  5.9& -18.5 $\pm$&  5.9&   N\\
 JCO6\_059 & 14.48& 1.27&  108 $\pm$&  -9999&  -14.9 $\pm$&   0.5&  20.0 $\pm$&    -9999&  -1.10&   0.21&   0.29&   -2.8 $\pm$&  5.9&  -6.9 $\pm$&  5.9&   N\\
 JCO6\_088 & 13.72& 1.03&  223 $\pm$&     19&  -13.2 $\pm$&   0.4&  20.0 $\pm$&    -9999&  -0.50&   0.29&   0.31&    0.0 $\pm$&  0.0&   0.0 $\pm$&  0.0&   Y\\
 JCO6\_095 & 13.51& 1.01&  285 $\pm$&     46&  -13.7 $\pm$&   0.9&  17.9 $\pm$&      2.9&  -0.43&   0.28&   0.42&    3.2 $\pm$&  5.9&  -0.9 $\pm$&  5.9&   Y\\
 JCO6\_111 & 15.99& 1.78&   64 $\pm$&  -9999&  -14.0 $\pm$&   0.4&  20.0 $\pm$&    -9999&   0.59&   0.68&   0.63&    7.0 $\pm$&  6.0&  -7.8 $\pm$&  5.9&   Y\\
 JCO6\_240 & 11.66& 0.59&   72 $\pm$&  -9999&  -11.9 $\pm$&   4.3&  75.9 $\pm$&      4.4&  -3.16&   0.23&   0.24&   -4.7 $\pm$&  1.5&  -8.5 $\pm$&  1.5&   Y\\
 JCO7\_021 & 13.13& 1.02&  241 $\pm$&     16&  -13.9 $\pm$&   0.8&  24.6 $\pm$&      2.0&  -0.40&   0.27&   0.34&    1.8 $\pm$&  5.9&  -4.7 $\pm$&  5.9&   Y\\
 JCO7\_027 & 14.20& 1.22&  267 $\pm$&  -9999&  -11.7 $\pm$&   1.2&  20.0 $\pm$&    -9999&  -1.14&   0.17&   0.24&   -5.5 $\pm$&  5.9&  -0.8 $\pm$&  5.9&   N\\
 JCO7\_079 & 13.09& 0.85&   83 $\pm$&     24&  -13.5 $\pm$&   1.1&  20.0 $\pm$&    -9999&  -1.57&   0.22&   0.23&    2.0 $\pm$&  2.8&  -2.5 $\pm$&  2.6&   Y\\
 JCO7\_088 & 15.41& 1.88&  117 $\pm$&  -9999&  -12.9 $\pm$&   0.9&  24.0 $\pm$&      3.5&   0.69&   0.81&   0.73&   -4.0 $\pm$&  6.4&  -3.2 $\pm$&  5.9&   Y\\
 JCO7\_130 & 14.14& 1.32&   93 $\pm$&  -9999&  -12.2 $\pm$&   0.7&  20.0 $\pm$&    -9999&  -1.16&   0.22&   0.35&   -7.0 $\pm$&  5.9&   2.2 $\pm$&  5.9&   N\\
 JCO7\_670 & 17.35& 2.58&  157 $\pm$&  -9999&  -11.1 $\pm$&   7.3&  35.2 $\pm$&      5.2&   1.02&   0.64&   0.66&             &     &            &     &   Y\\
 JCO8\_061 & 14.45& 1.28&   79 $\pm$&  -9999&  -14.0 $\pm$&   0.6&  20.0 $\pm$&    -9999&  -1.10&   0.23&   0.26&    1.4 $\pm$&  5.9& -16.8 $\pm$&  5.9&   N\\
 JCO8\_257 & 12.92& 1.05&  132 $\pm$&     30&  -13.5 $\pm$&   1.0&  28.8 $\pm$&      2.2&  -1.72&   0.24&   0.26&   -5.2 $\pm$&  3.0&  -6.1 $\pm$&  2.9&   Y\\
 JCO8\_364 & 17.53& 2.62&  174 $\pm$&  -9999&  -15.2 $\pm$&   1.4&  23.7 $\pm$&      4.1&   1.62&   0.71&   0.66&             &     &            &     &   Y\\
 JCO8\_395 & 17.65& 2.52&  241 $\pm$&  -9999&  -15.4 $\pm$&   2.1&  20.0 $\pm$&    -9999&   0.76&   0.77&   0.67&             &     &            &     &   Y\\
 JCO8\_550 & 17.87& 2.84&  480 $\pm$&  -9999&  -12.1 $\pm$&   5.9&  46.1 $\pm$&      4.7&   3.58&   0.91&   0.62&             &     &            &     &   Y\\
 JCO9\_120 & 14.40& 1.10&   94 $\pm$&  -9999&  -13.2 $\pm$&   0.5&  20.0 $\pm$&    -9999&  -1.17&   0.27&   0.30&    4.9 $\pm$&  5.9&   1.9 $\pm$&  5.9&   Y\\
 JCO9\_281 & 14.96& 1.42&  107 $\pm$&     19&  -13.4 $\pm$&   0.4&  20.0 $\pm$&    -9999&  -0.22&   0.45&   0.49&   -3.2 $\pm$&  6.1&  -6.9 $\pm$&  5.9&   Y\\
 JCO9\_406 & 14.54& 1.21&  211 $\pm$&  -9999&  -21.3 $\pm$&   6.1&  25.0 $\pm$&    -9999&  -1.00&   0.28&   0.28&   -3.1 $\pm$&  5.9&  -2.2 $\pm$&  5.9&   N\\
 JCO9\_508 & 13.51& 1.27&   58 $\pm$&  -9999&  -14.8 $\pm$&   0.4&  20.0 $\pm$&    -9999&  -1.13&   0.20&   0.27&   -0.5 $\pm$&  5.9&   1.2 $\pm$&  5.9&   N\\
\hline
\end{tabular}
\label{tab2}
\end{table*}

Cluster membership was assessed primarily from the RVs. The cluster is
clearly present as a concentration in the RV distribution (see
Fig.~\ref{lirv}), although it sits upon a much broader contaminating
field star distribution.  The subset of stars with strong Li~6708\AA\
features have very similar RVs. Discarding several outliers, there are
22 Li-rich stars, which we assume are cluster members, with
$-15.2<$RV$<-10.1$\,km\,s$^{-1}$, giving a weighted mean RV of
$-12.9\pm 0.3$\,km\,s$^{-1}$. Using spectra of all the observed RV
standards we estimate an additional external error on this mean of $\pm
0.5$\,km\,s$^{-1}$, making our value consistent with the $-12.9\pm
0.4$\,km\,s$^{-1}$ found by PG94, but $3.0\pm 1.2$\,km\,s$^{-1}$ higher
than the value reported by M08.  Comparing the standard deviation of
the RVs with their individual uncertainties indicates an intrinsic
cluster RV dispersion of 1\,km\,s$^{-1}$. Adding this internal
dispersion to the RV uncertainties
in quadrature, we then accept any star with an RV within 1.5-sigma of
the cluster mean as a probable cluster member, yielding 56
candidates. Applying similar criteria for assumed mean RVs
10\,km~s$^{-1}$ either side of the IC~4665 RV allows us to estimate
that about $16\pm 3$ of these 56 candidates are non-members with RVs
that randomly coincide with IC~4665.

The candidate list was refined in several ways. Figure~1 shows that
the heaviest contamination is expected for $1.0<V-I<1.5$ where a ``finger'' of
background K-giants cuts across the cluster sequence. To filter
these out, we measured the EWs of two gravity-sensitive, isolated Ca\,{\sc
i} lines at 6439\AA\ and 6463\AA. We have used the spectral synthesis
software {\sc uclsyn} (Smalley, Smith \& Dworetsky 2001) along with
Atlas stellar atmosphere models (Kurucz 1993) to establish that these
should be a about a factor of 2 weaker in giants than in dwarfs of the
same temperature/colour. Accordingly, in Fig.~\ref{cavi} we identify 10
of our candidates, all with $1.2<V-I<1.33$, that have weaker Ca\,{\sc
i} lines than expected for dwarfs and which we class as probable giants
(or dwarfs with a low metallicity) and hence not cluster members. As
expected, none of these have measurable Li, rapid rotation or H$\alpha$
emission (see below).

Proper motions for 45 of the 56 RV candididates are found in the {\sc nomad}
database (Zacharias et al. 2004). 
The mean proper-motion of the Li-rich stars is $-0.7 \pm 1.0$
milli-arcsec/yr in RA and $-6.2\pm 0.8$ milli-arcsec/yr in Dec. This is
not greatly different from the mean for field stars in this
direction. We find that 2 stars (the brightest F-star in our sample
[JCO4\_295] and an M-dwarf with H$\alpha$ in absorption, [JCO4\_558])
have proper-motions incompatible with cluster membership at the 3-sigma
level.

We measured the EW of the H$\alpha$ line by direct integration above
(or below) a pseudo-continuum. Most of the cooler targets exhibit
H$\alpha$ emission (see Fig.~\ref{cavi}), as expected for young,
chromospherically active stars. It is unlikely that the 4 cool
targets ($V-I>2$) with H$\alpha$ in absorption are members, because
H$\alpha$ emission is ubiquitous for such stars, even in older clusters
like IC~2391 and the Pleiades (see Stauffer et al. 1997).  These four
objects are {\it probably} older contaminating field stars and one of
them is also a non-member by proper motion as explained above. We wish
to use H$\alpha$ emission as an age indicator later on (see
section~\ref{discuss}), so we do not discard these objects, but mark 
them as ``possible'' members.

Finally, we have taken $JHK$ magnitudes from 2MASS (Cutri et al. 2003)
and plotted the $J-H$ vs $H-K$ diagram. Only one star stands out from
the general trend of photospheric colours, and this only by
2.5-sigma. JCO5\_329 is a target with $J-H=0.15\pm0.04$ and
$H-K=0.15\pm0.04$, which is too red in $H-K$ by about $0.10$\,mag
compared with other cluster candidates. This could perhaps be due to
warm circumstellar dust, although this is unlikely at an age of $\simeq
30$\,Myr. For now this star is classed as a ``possible'' member.

The 40 stars that pass all these tests are classified as definite
members. The full 56-star candidate list and a summary of the
measurements and membership judgements are shown in Fig.~\ref{vvi2} and
listed in Table~2.

\subsection{Comparison with other work on IC~4665}

Many of our targets and members were selected
as such and studied by Prosser (1993), PG94, MM97
and M08.  We comment in detail on the comparison of our measurements and
membership selections with these papers below.

We find 137 of our targets have $V$ and $I$ photometry in Prosser
(1993). The agreement of this photometry with that presented here is
good. The average difference in $V$ magnitudes (ours minus Prosser) is
$-0.028 \pm 0.007$ mag with a standard deviation of 0.086 mag. After
converting Prosser's $V-I$ Kron colours to $V-I$ Cousins colours using
the formulae of Bessell \& Weis (1987), the average difference in
colour is $-0.033 \pm 0.007$ mag with a standard deviation of 0.077
mag and no obvious colour dependence in the range $0.5<V-I<3$.

There are 30 stars with RVs common to our work and PG94. The agreement
is excellent apart from 4 objects. Three of these (JCO4\_057$=$P166 ,
JCO5\_160$=$P19 and JCO8\_038$=$P38) have been identified by us as
probable SB2 binary systems (see Table~1).  The other (JCO9\_462$=$P66)
has a RV discrepancy (our value minus that of PG94) of $-17.2 \pm
0.5$\,\kms\ and it is probably a short period SB1 binary.  For the
remaining 26 stars the average RV discrepancy is $-0.25 \pm
0.41$\,\kms\ with a standard deviation of 2.1\,\kms. Ignoring the SB2
objects mentioned above, then the $v\sin i$ values quoted by PG94 also
agree well with our measurements. The one exception is JCO6\_347$=$P12,
where PG94 find $v \sin i <10$\,\kms\ but we have $v \sin i = 108\pm
4$\,\kms. As both MM97 and Allain et al. (1996) also find that P12 is a
rapid rotator, with $v\sin i =70$\,\kms\ and a rotation period of 0.60~days
respectively, then it seems likely that the PG94 measurement is in
error. We have 12 $v\sin i$ measurements for targets in common with
Shen et al. (2005) and all are in perfect agreement.

There are 59 targets common to this paper and M08. A comparison of RVs
reveals three probable SB1 binaries (JCO5\_039 $=$ number 103 in M08,
with RV difference of $24.2\pm 0.8$\,\kms; JCO4\_271 $=$ M08 number 40,
$\Delta$RV$=9.2\pm 1.0$\,\kms; JCO4\_249 $=$ M08 number 116,
$\Delta$RV$=9.9\pm1.0$\,\kms). After clipping out these objects, the
mean difference between our RVs and those of M08 is $2.9\pm 0.3$\,\kms,
with a standard deviation of 1.9\,\kms.

We have been able to compare our values for EW[Li] with 12 stars also
studied by MM97. Agreement is good with the average EW difference (ours
minus MM97) being $-18 \pm 9$\,m\AA\ with a standard deviation of
27\,m\AA.  We also have 5 targets with EW[Li] listed by M08 and again,
the agreement is good.

In terms of selected members, there are 7 objects selected by PG94 or
by MM97 as cluster members that are not in our list of candidate
members (Table~2). For P12, P19, P27, P38 and P166 these either have an
RV outside our selection range or are possible SB2 binary systems. The
objects P39 and P155 may be genuine members but are among the 13 per
cent of photometric candidates we did not observe spectroscopically.
There are 27 candidate members considered by M08 that do not appear in
our Table~2. Most of these are too faint ($V>18$) to have been included
in our sample, but six (identified by M08 with the numbers 6, 16, 20, 59, 83
and 125) were observed by us but had RV values outside our
selection range. JCO5\_190 ($=$ M08 number 59) is a probable binary
system according to both us and M08. JCO6\_038 ($=$ M08 number 20) is a
binary according to M08, but we have only one RV measurement.  M08
claim weak detections of Li with EWs between 30 and 70\,m\AA\ in stars
6, 16, 20, 59 and 126. Our measurements are in agreement apart from for
star 6 ($=$ JCO6\_063) where we find EW[Li]$=99\pm 31$\,m\AA\ compared
with the $52\pm 5$\,m\AA\ measured by M08.

\section{Lithium Depletion in IC~4665}

\begin{figure*}
\centering
\begin{minipage}[t]{0.48\textwidth}
\includegraphics[width=82mm]{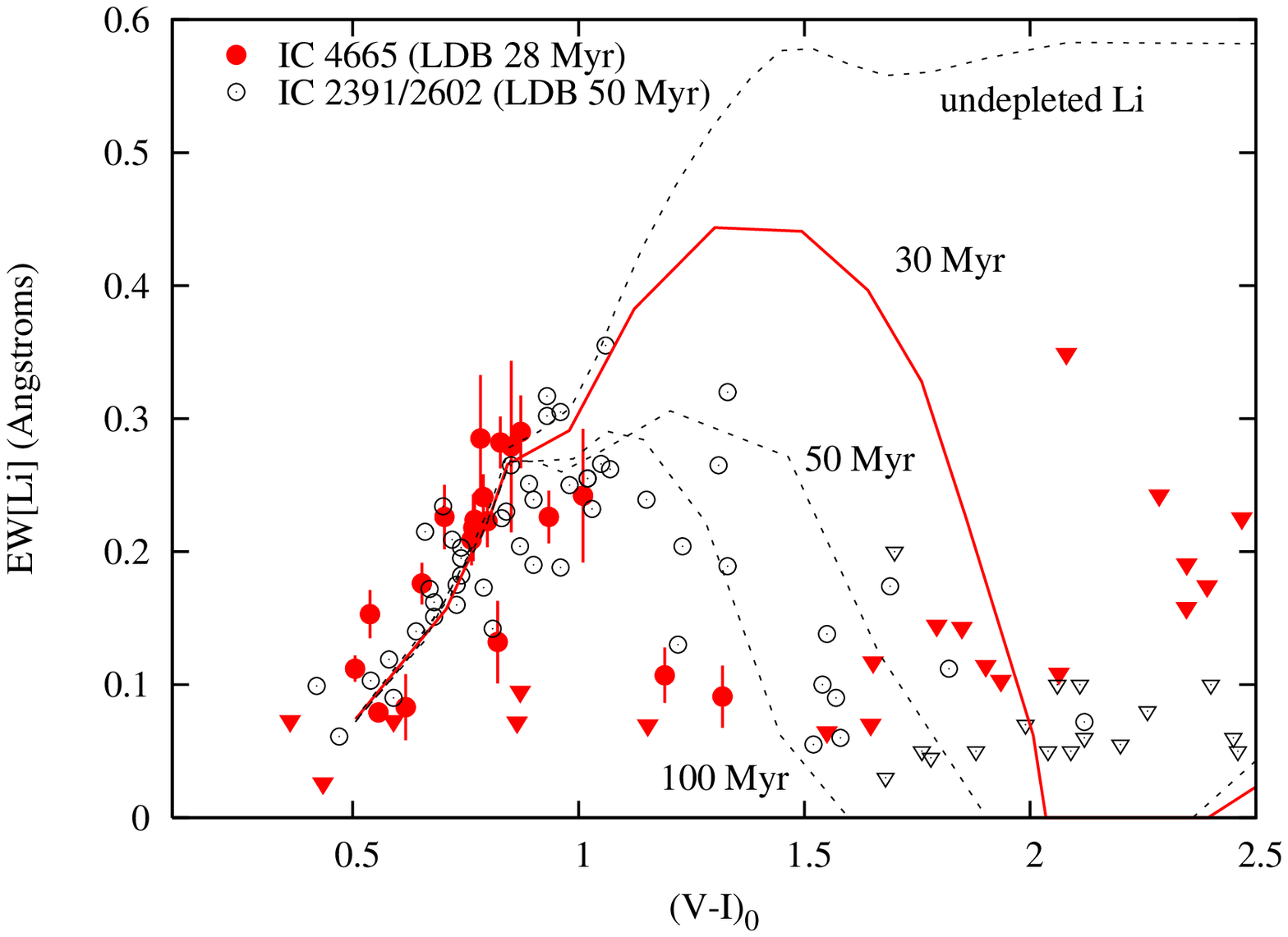}
\end{minipage}
\begin{minipage}[t]{0.48\textwidth}
\includegraphics[width=82mm]{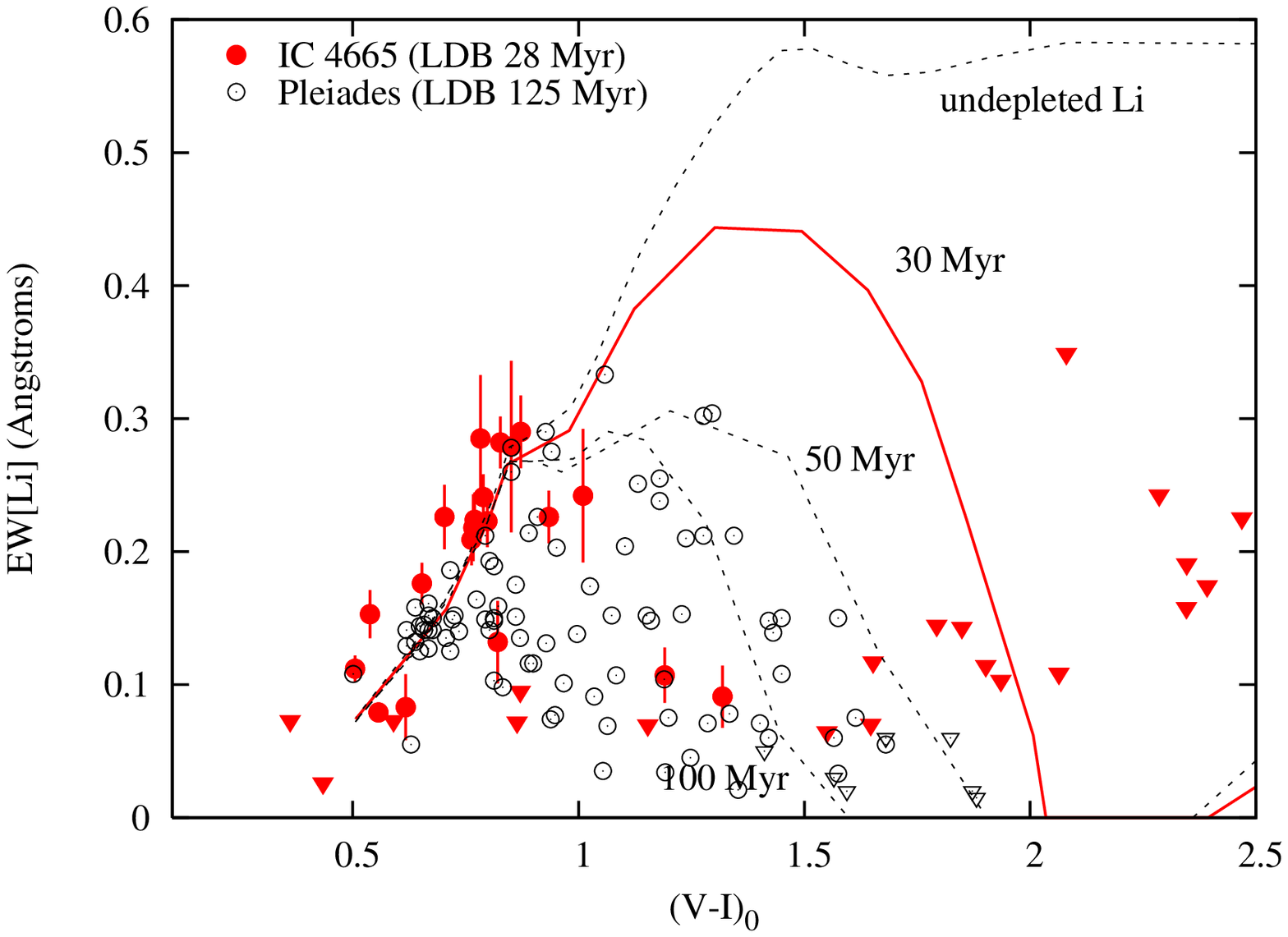}
\end{minipage}
\caption{The EW of the Li\,{\sc i}~6708\AA\ feature versus intrinsic
  $V-I$ for the IC~4665 candidate members (filled symbols, LDB age
  $28\pm 5$\,Myr) 
  compared with those in (a) the IC~2391 and IC~2602 clusters (open
  symbols, with an LDB age of $50\pm 5$\,Myr)
  and (b) the Pleiades cluster (open symbols, with an LDB age of
  $125\pm 8$\,Myr). In both plots triangles indicate upper limits. Also
  shown are model isochrones of Li depletion derived from the Baraffe
  et al. (2002) evolutionary models. Despite having a significantly
  younger LDB age, stars with $1<(V-I)_0<2$ in IC~4665 seem
  significantly more depleted than the majority of objects in the older
  IC~2391/2602 clusters, more similar to the bulk of objects in the
  much older Pleiades and way below the predicted depletion levels for a
  30\,Myr old population.
}
\label{liviplot}
\end{figure*}

\subsection{Lithium Equivalent Widths}

Figure~\ref{liviplot} compares EW[Li] as a function of intrinsic $V-I$
in IC~4665 (from Table~2 excluding the 12 non-members, so only a few
contaminants may be left and these may have already been identified as
the ``possible'' members -- see section~\ref{membership}) with EW[Li]
measured in a similar way for stars in the IC~2391, IC~2602 and
Pleiades clusters. The contribution of the weak, blended Fe\,{\sc
i}~6707.44\AA\ line has not been removed from any of the EWs at this
stage.  We have initially chosen to make the comparison in this
observational plane because the relative uncertainties are easier to
understand and besides which, as the clusters have similar metallicity
(see section~\ref{discuss}) the transformation to Li abundance versus
$T_{\rm eff}$ is similar for each of these clusters.

IC~2391 has a measured LDB age of $50\pm 5$\,Myr (Barrado y Navascues
et al. 2004) and IC~2602 is usually assumed to have a similar age based
on low-mass isochronal fits.  The Li data come from the analyses of
Randich et al. (1997) and 
Randich et al. (2001), the photometry from papers by Patten \& Simon
(1996) and Prosser, Randich \& Stauffer (1996).  The colours of the
plotted data have been corrected for reddening ($E(V-I)=0.01$ and 0.04
for IC~2391 and IC~2602 respectively -- Patten \& Simon 1996) . Our
purpose is to show that whilst the amount of Li-depletion seen among F-
and G-type stars ($0.4<(V-I)_0<0.9$) is quite similar in IC~4665 and
IC~2391/2602 (actually, it is consistent with very little depletion at
all, see below and section~\ref{abundances}), 
the few K- and early M-type stars we have found in IC~4665 seem to have
undergone as much {\it or more} Li-depletion, despite this cluster having a
younger LDB age. Note that uncertainties in the relative cluster reddenings
(dominated by the $\pm 0.06$\,mag in $E(V-I)$ for IC~4665) 
affect the comparison of F/G stars far more than for cooler stars.

The status of the K- and M-stars becomes more significant when
considered in the context of the Li depletion predicted by PMS models.
Figure~\ref{liviplot} also shows loci of EW[Li] expected for undepleted
lithium, along with isochrones at ages of 30, 50 and 100\,Myr,
determined using the Baraffe et al. (2002) evolutionary models with
convective mixing length set to 1.0 pressure scale height.  These
isochrones were calculated using a relationship between
$(V-I)_0$ and effective temperature from Kenyon \& Hartmann (1995), a
curve of growth for the Li\,{\sc i}~6708\AA\ feature described in
Jeffries et al. (2003) and an assumed initial Li abundance of
$A$(Li)$=3.3$, on the conventional logarithmic scale (Anders \&
Grevesse 1989).  The curve
representing an undepleted level of Li is an excellent match to EW[Li]
measurements for cool objects in star forming regions over the whole
colour range considered (e.g. see Jeffries et al. 2009). The lower
EW[Li] in stars of the same colour in young PMS clusters like IC
2391/2602 has long been cited as evidence for PMS Li depletion
(e.g. Randich et al. 2001).

Evolutionary models by different authors predict different levels of Li
depletion as a function of time (e.g. Pinsonneault, Kawaler \& Demarque
1990; Pinsonneault 1997; D'Antona \& Mazzitelli 1997; Siess, Dufour \&
Forestini 2000; Piau \& Turck-Chi\`eze 2002) . We have chosen the
Baraffe et al. (2002) models because they reasonably match the average
Li depletion seen in IC 2391/2602 at their LDB age, although
underpredict the Li-depletion seen in some of the cooler stars.
However, if IC~4665 were $\simeq 20$\,Myr younger than IC~2391/2602, as
suggested by its LDB age, we would expect to see stars with much higher
EW[Li] (at the same intrinsic colour) than in IC~2391/2602 for
$(V-I)_0>1$, and to have detected lithium in stars as cool as $(V-I)_0
\simeq 2$. Whether or not the detailed predictions of these models are
correct, {\it all} published models predict a significant diminuition
of EW[Li] with age at the colour of K- and M-stars. The observed
progression of EW[Li] between the K- and M-stars ($1.0<(V-I)_0 <2.0$)
of IC~4665 and IC~2391/2602 is, if anything, the wrong way around.

Figure~\ref{liviplot}b shows a similar comparison with Li data from the
Pleiades, which has an LDB age of $125\pm 8$\,Myr (Stauffer et
al. 1998). The Li EWs come from the uniform re-analysis by Sestito et
al. (2005); upper limits for cool stars were added from Garc\'ia-L\'opez,
Rebolo \& Mart\'in (1994) and Jones et
al. (1996).  Kron $V-I$ photometry from Stauffer (1982, 1984) and
Prosser, Stauffer \& Kraft (1991) has been converted to the Johnson-Cousins
system using the transformation in Bessell \& Weis (1987). A reddening
of $E(V-I)=0.05$ was assumed, except for those few stars listed in Table~9 of
Soderblom et al. (1993a), which have larger reddening values. 
The F- and G-stars of the Pleiades have
marginally lower EW[Li] than their equivalents in IC~4665. This may
indicate some Li-depletion, but the uncertainty in $E(V-I)$ for IC~4665
could eliminate most of this difference.  The EW[Li] of the K- and
M-stars of IC~4665 appear consistent with the most Li-depleted K- and
M-stars in the Pleiades.  Despite being much older, a significant
fraction of the Pleiades population lies above the 
existing data in IC~4665.

We have checked through all of our IC~4665 spectra to see whether any
possible cluster members with high levels of Li might have been
excluded during the RV selection process. There are a number of objects
with significant Li\,{\sc i}~6708\AA\ detections that are either young
field stars or candidate short-period binary stars in IC~4665 (see
Fig.~\ref{lirv} and Table~1). However, none of these lie above the
envelope of IC~4665 candidate members in Fig.~\ref{liviplot}, so their
inclusion could only reinforce the phenomena described above.

\subsection{Lithium Abundances}
\label{abundances}

\begin{figure*}
\centering
\begin{minipage}[t]{0.48\textwidth}
\includegraphics[width=82mm]{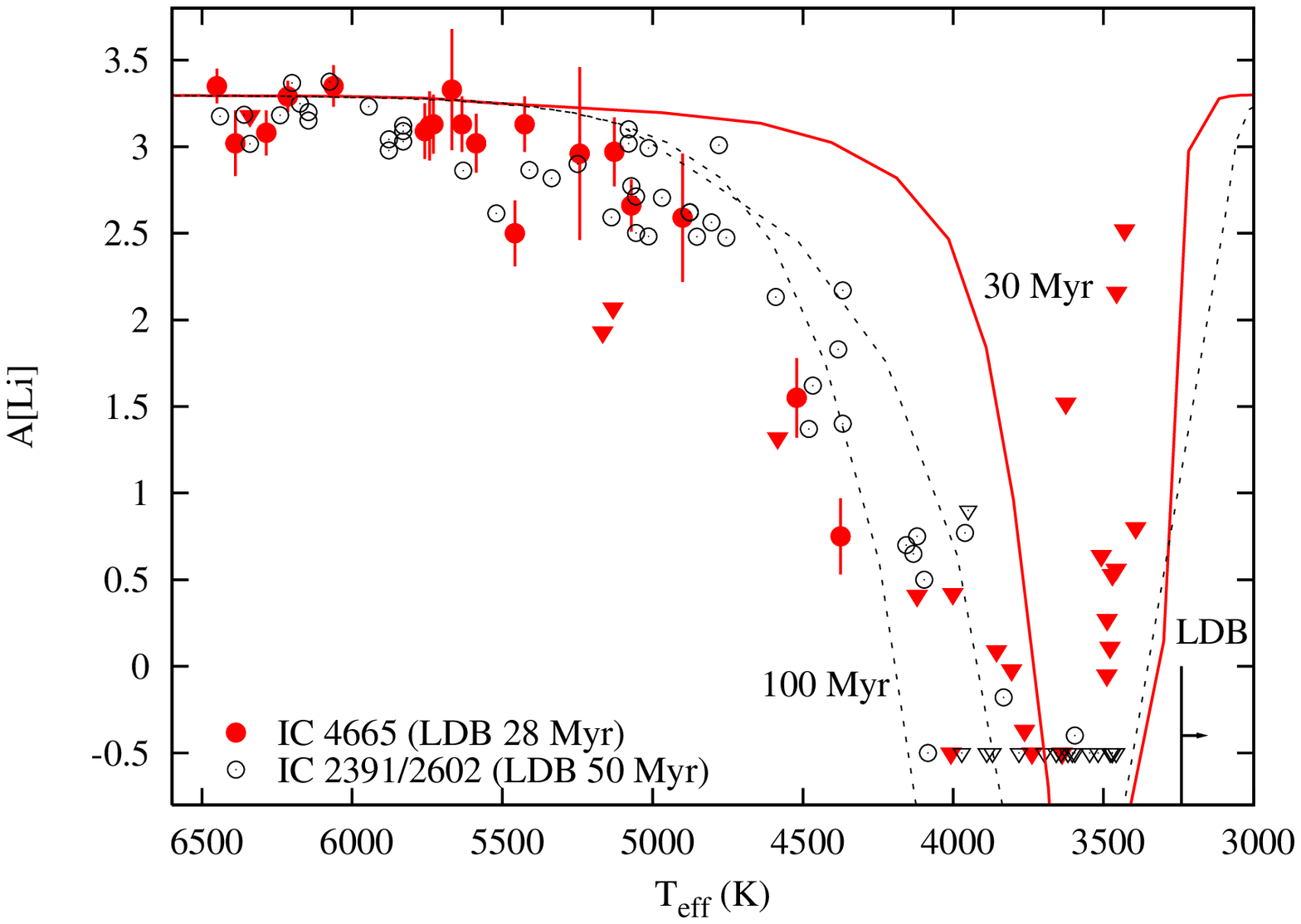}
\end{minipage}
\begin{minipage}[t]{0.48\textwidth}
\includegraphics[width=82mm]{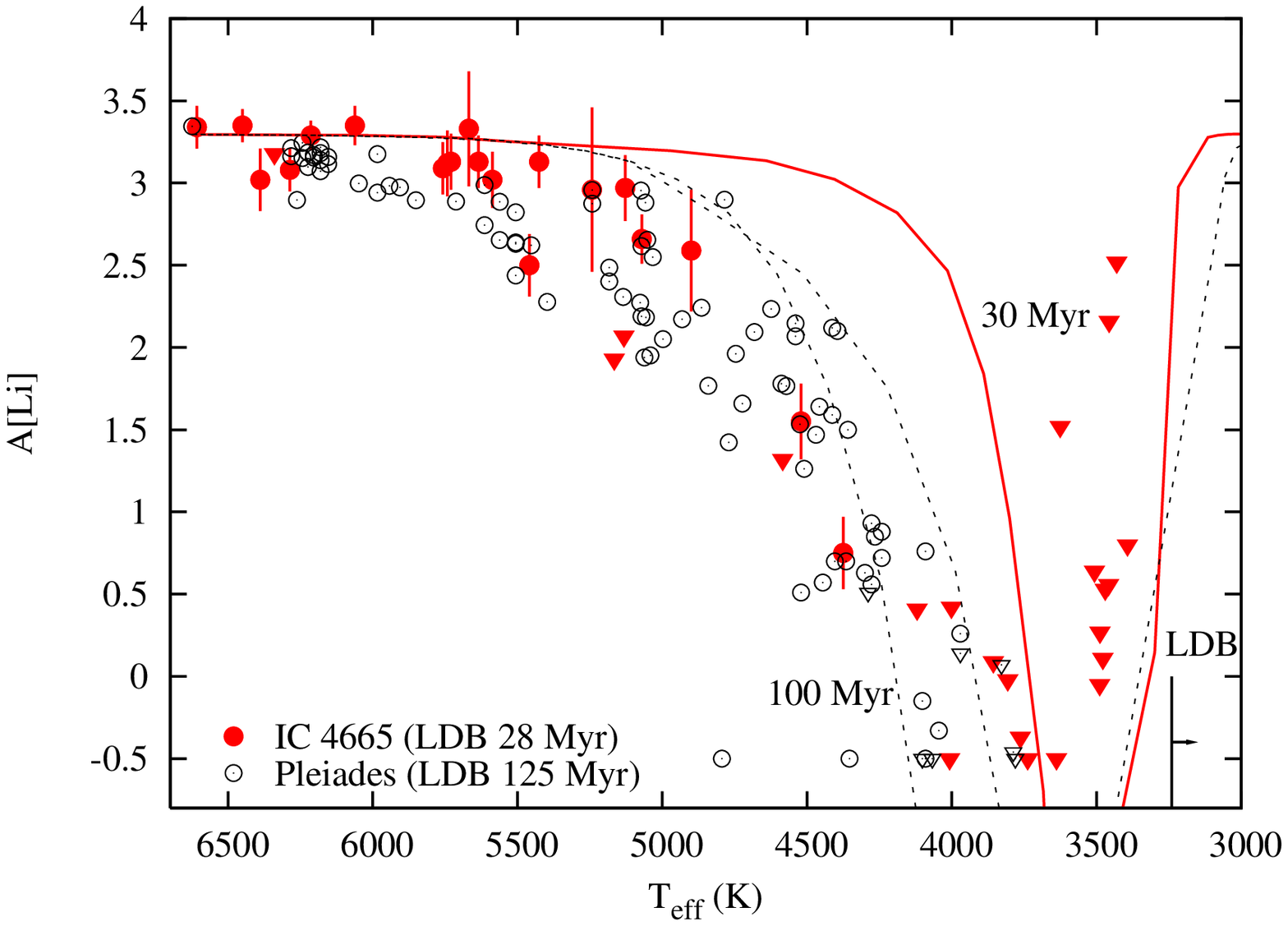}
\end{minipage}
\caption{A comparison of lithium abundances in IC~4665 (filled symbols)
  with those in (a) the IC~2391 and IC~2602 clusters (open
  symbols)
  and (b) the Pleiades cluster (open symbols). 
  In both plots triangles indicate upper limits. Also
  shown are model isochrones of Li depletion from the Baraffe
  et al. (2002) evolutionary models at ages of 30, 50 and
  100\,Myr. The vertical bar at 3240\,K indicates the approximate
  $T_{\rm eff}$ of the LDB for IC~4665 found by M08.}
\label{abunplot}
\end{figure*}

\begin{table*}
\caption{Lithium abundances in IC~4665. Columns list the star, a $V-I$,
  the calculated $T_{\rm eff}$, the LTE Li abundance ($A$(Li) = 12 +
  $\log [N$(Li)/$N$(H)$]$) and its uncertainty and an NLTE Li abundance for stars with
  $T_{\rm eff}>4500$\,K (see text).}
\begin{tabular}{cccrrr|cccrrr}
\hline
Identifier & $V-I$ & $T_{\rm eff}$ (K)& $A$(Li) &
 $\sigma A$(Li)&$A$(Li) & Identifier & $V-I$ & $T_{\rm eff}$ (K)& $A$(Li) &$\sigma A$(Li)&$A$(Li)\\
   & & (K) & LTE &  &NLTE & & & (K) & LTE & & NLTE \\
\hline
 JCO1\_427 &   0.997 & 5742 &      3.33 &  0.20 &     3.12&   JCO5\_280 &   1.094 &5166 & $<$  1.82 &     &$<$ 1.93 \\
 JCO1\_530 &   0.787 & 6389 &      3.12 &  0.19 &     3.02&   JCO5\_282 &   0.735 &6608 &      3.51 & 0.13&    3.34 \\
 JCO2\_145 &   0.768 & 6450 &      3.54 &  0.10 &     3.35&   JCO5\_296 &   1.081 &5243 &      3.17 & 0.50&    2.96 \\
 JCO2\_213 &   0.933 & 6062 &      3.64 &  0.12 &     3.35&   JCO5\_329 &   0.664 &7099 &           &     &         \\
 JCO2\_220 &   2.023 & 3856 & $<$  0.09 &       &         &   JCO5\_394 &   2.132 &3763 & $<$ -0.37 &     &         \\
 JCO2\_373 &   0.883 & 6214 &      3.50 &  0.09 &     3.29&   JCO5\_472 &   2.165 &3738 & $<$ -0.50 &     &         \\
 JCO2\_637 &   2.745 & 3456 & $<$  2.16 &       &         &   JCO5\_515 &   2.741 &3458 & $<$  0.56 &     &         \\
 JCO2\_736 &   2.986 & 3393 & $<$  0.80 &       &         &   JCO5\_521 &   2.294 &3638 & $<$ -0.50 &     &         \\
 JCO3\_065 &   1.549 & 4375 &      0.75 &  0.22 &         &   JCO6\_088 &   1.028 &5587 &      3.20 & 0.17&    3.02 \\
 JCO3\_285 &   0.820 & 6340 & $<$  3.33 &       & $<$ 3.18&   JCO6\_095 &   1.013 &5668 &      3.71 & 0.35&    3.33 \\
 JCO3\_357 &   1.102 & 5128 &      3.20 &  0.20 &     2.97&   JCO6\_111 &   1.780 &4120 & $<$  0.41 &     &         \\
 JCO3\_395 &   1.057 & 5426 &      3.41 &  0.16 &     3.13&   JCO6\_240 &   0.591 &7418 &           &     &         \\
 JCO3\_396 &   0.993 & 5758 &      3.29 &  0.16 &     3.09&   JCO7\_021 &   1.019 &5635 &      3.37 & 0.16&    3.13 \\
 JCO3\_724 &   2.310 & 3625 & $<$  1.52 &       &         &   JCO7\_079 &   0.847 &6286 &      3.20 & 0.13&    3.08 \\
 JCO3\_770 &   2.699 & 3470 & $<$  0.53 &       &         &   JCO7\_088 &   1.882 &4001 & $<$  0.42 &     &         \\
 JCO4\_053 &   1.164 & 5071 &      2.73 &  0.15 &     2.66&   JCO7\_670 &   2.576 &3489 & $<$ -0.05 &     &         \\
 JCO4\_226 &   1.000 & 5730 &      3.36 &  0.17 &     3.13&   JCO8\_257 &   1.051 &5459 &      2.49 & 0.19&    2.50 \\
 JCO4\_337 &   1.383 & 4584 & $<$  1.06 &       & $<$ 1.32&   JCO8\_364 &   2.622 &3478 & $<$  0.11 &     &         \\
 JCO4\_437 &   2.079 & 3806 & $<$ -0.02 &       &         &   JCO8\_395 &   2.516 &3507 & $<$  0.64 &     &         \\
 JCO4\_459 &   2.577 & 3488 & $<$  0.27 &       &         &   JCO8\_550 &   2.839 &3430 & $<$  2.52 &     &         \\
 JCO4\_591 &   1.877 & 4007 & $<$ -0.50 &       &         &   JCO9\_120 &   1.101 &5132 & $<$  1.97 &     & $<$1.86 \\
 JCO5\_179 &   1.240 & 4900 &      2.63 &  0.37 &     2.59&   JCO9\_281 &   1.421 &4521 &      1.29 & 0.23&    1.55 \\
\hline
\end{tabular}
\label{liabundances}
\end{table*}

An alternative way to view these comparisons is to calculate Li
abundances. The procedure described in detail by Jeffries et
al. (2003) was used to convert intrinsic colours to $T_{\rm eff}$ and then
transform EW[Li] into an abundance using a curve of growth for the
Li\,{\sc i}~6708\AA\ feature. The curve of growth is 
based on the calculations presented by
Soderblom et al. (1993b) for $T_{\rm eff}>4500$\,K and those of
Zapatero-Osorio et al. (2002) for cooler objects. Prior to the
transformation, EW[Li] was corrected for the presence of the weak, blended
Fe\,{\sc i} line, using the formula EW[Fe]$=58 - 0.01 T_{\rm eff} +
3.8\times10^{-7} T_{\rm eff}^2$, derived from the
correction as a function of $B-V$ suggested by Soderblom et
al. (1993b) after applying the $B-V/T_{\rm eff}$ relation of Kenyon \&
Hartmann (1995). The curves of growth are based on LTE atmospheres.
NLTE corrections to the Li abundances 
were made using the results of Carlsson et
al. (1994).  The corrections vary from about $-0.2$\,dex for $T_{\rm eff}
= 6000$\,K and $A$(Li)$_{\rm LTE}=3.4$, to $+0.25$\,dex for $T_{\rm
eff}=4500$\,K and $A$(Li)$_{\rm LTE}=1.0$.  The Carlsson et al. (1994)
corrections are only calculated for $T_{\rm eff}>4500$\,K, so for cooler
stars no corrections have been applied. For $T_{\rm eff} \leq
4000$\,K it is likely that the absolute corrections become smaller
(Pavelenko \& Magazzu 1996).
 
The Li abundances for the IC~4665 members (and possible members) from
Table~2 are given in Table~\ref{liabundances}. Note that any abundances
$A$(Li)$<-0.5$ are extrapolations beyond the calibrated limits of the
curves of growth and are pegged at this value. Similarly, we have not
derived abundances for stars with $T_{\rm eff}>6700$\,K.  The abundance
uncertainties are derived from propagated uncertainties in EW[Li]
combined in quadrature with errors due to random $T_{\rm eff}$
uncertainties, assumed to be $\pm 100$\,K.  There are also a
number of systematic uncertainties to consider which are discussed at
length by Jeffries et al. (2003). Most of these, such as the adopted
temperature scale and curves
of growth, affect stars in all the clusters considered and do not
alter comparisons between them. However, there are systematic
uncertainties at the $\pm 0.2$\,dex level which will affect the
comparison with models of Li depletion and the assumed 
initial Li abundance also has some uncertainty. We used an initial
$A$(Li)$=3.3$, on the basis of the meteoritic Li abundance in the
solar system (Anders \& Grevesse 1989). 
It seems reasonable to assume that the initial abundance
is similar for all the clusters considered, because they have very
similar metallicities (see section~\ref{discuss}), but there could
still be systematic errors of $\sim 0.1$\,dex in the absolute
model abundances.

Figure~\ref{abunplot} shows comparisons between IC~4665,
IC~2391/2602, the Pleiades and the Baraffe et al. (2002) Li depletion
models in the $A$(Li)-$T_{\rm eff}$ plane. Because the transformations
are similar for each cluster then the points made in the last
subsection by comparing EW[Li] are equally valid here -- namely, that
the cool IC~4665 members are as or more depleted than stars of the same
$T_{\rm eff}$ in IC~2391/2602 and occupy the same region as the most
Li-depleted stars in the Pleiades. There is some evidence that the
hotter stars ($T_{\rm eff}>5000$\,K) in IC~4665 have depleted a little 
less Li than their counterparts in the Pleiades, 
but again we caution that this conclusion
rests on the adopted reddening for IC~4665. A reduction in $E(V-I)$
of 0.06 mag, would make the hotter stars cooler by about 100\,K and
reduce $A$(Li) by about 0.15\,dex. At lower temperatures 
($T_{\rm eff}<5000$\,K) the assumed
reddening has less influence; any uncertainty shifts the
points almost parallel to the trend of decreasing abundance with
decreasing temperature seen in the data and models, 
and hence does not greatly alter the comparison between one
cluster and another or the comparison with the models.

The comparison with the models shows again that IC~4665 appears to
have suffered much more depletion than expected. Other models with
greater convective efficiency (e.g. the Baraffe models with a mixing
length of 1.9 pressure scale heights) could address this, but these
models would also predict far too much depletion to account for
the observed Li abundances in IC~2391/2602 and the Pleiades. 

The $T_{\rm eff}$ below which Li re-appears on the other side of the ``Li-chasm''
can be estimated from the data in M08. The LDB
occurs at $I-K \simeq 2.7$, corresponding to $T_{\rm eff} \simeq
3240\,K$ from the colour-$T_{\rm eff}$ relations of Kenyon \& Hartmann
(1995). This appears to be in perfect agreement with the predictions of
the 30\,Myr Baraffe et al. (2002) isochrone, though it should be noted
that there are probably uncertainties of 100-200\,K in our estimated LDB $T_{\rm eff}$.

\section{Discussion}
\label{discuss}

\subsection{The age of IC~4665}

\begin{figure}
\centering
\includegraphics[width=80mm]{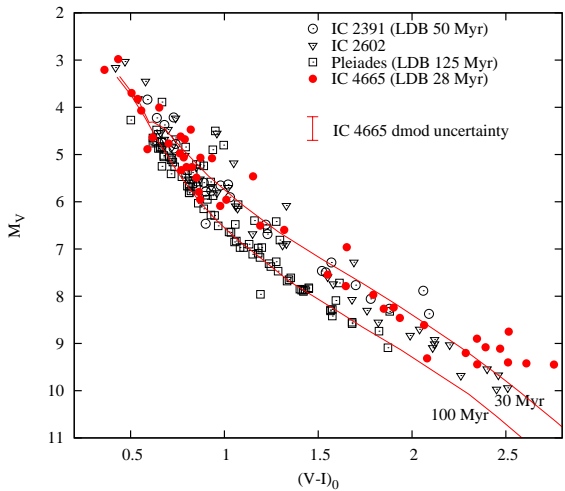}
\caption{An absolute $V$ vs intrinsic $V-I$ diagram for members of the
  IC~4665, IC~2391, IC~2602 and Pleiades clusters. Also shown are two
  model isochrones at 30 and 100\,Myr from Baraffe et al. (2002). The
  undertainty on the distance modulus of IC~4665 is indicated. The
  distance modulus uncertainties for the other clusters are small 
  in comparison. IC~4665 is definitely younger than the Pleiades in
  this diagram but could be as old or younger than IC~2391/2602.
}
\label{relativevvi}
\end{figure}

The LDB age of IC~4665 quoted by M08 is $28\pm 5$\,Myr, where the bulk
of the uncertainty is due to uncertainty in the distance ($370\pm
50$\,pc) to the cluster. The re-analysis of the Hipparcos data by van
Leeuwen (2007) improves matters only a little, where a mean distance to
IC~4665 of $356^{+35}_{-34}$\,pc is found. The LDB age is found from
the luminosity of the brightest very low-mass stars at temperatures below the
``Li-chasm'' (see Fig.~7)  which still retain
their initial lithium (see M08). Although age spreads of up to 10\,Myr
have been suggested in young clusters (Palla \& Stahler 2000; Jeffries
2007), there is no strong evidence in M08 for Li-depleted cluster members
below the LDB luminosity that might indicate the presence of a
significant fraction of older stars in the cluster. In any case, a
10\,Myr age spread would not alter any of our subsequent discussion and
conclusions.

The LDB age has
two key advantages over other age estimation techniques. First,
although it is dependent on an assumed distance, it is much less
sensitive to distance than some other methods, notably fitting
isochrones to either the high- or low-mass stars (see below). Second,
the physics of Li-depletion in fully convective low-mass stars is
thought to be well enough understood that little room for systematic
error remains. Indeed, any errors that persist must be due to effects
that are currently not being debated by the various ``flavours'' of
current low-mass evolutionary models, as they all give the same LDB
ages to within a few per cent for clusters at $\sim 30$\,Myr (see Burke
et al. 2004; Jeffries \& Oliveira 2005). For these reasons we attribute
the greatest confidence to both the absolute and relative ages of
clusters determined using the LDB technique.

We can now ask whether the absolute ages and relative age ordering of
clusters derived using other techniques concur with the LDB
ages?  For IC~4665, the age we would deduce from the low-mass
isochrones in Fig.~\ref{vvi2} (given a $\pm 0.25$\,mag distance modulus
uncertainty) is in the range 20--50\,Myr, probably younger than the
IC~2391 and IC~2602 clusters (see also M08) and certainly younger
than the Pleiades.  These absolute isochronal ages could be in error so
it helps to look at relative ages. Figure~\ref{relativevvi} plots
absolute $V$ versus intrinsic $V-I$ diagram for these clusters,
using the points plotted in Fig.~\ref{liviplot} and using intrinsic
distance moduli of 6.05\,mag, 5.95\,mag and 5.54\,mag for IC~2391, IC~2602
and the Pleiades respectively (Stauffer et al. 1997).

From Fig.~\ref{relativevvi} we can conclude that IC~4665 is younger
than the Pleiades, but depending on the exact distance (the uncertainty
is indicated in the diagram) it could be younger than IC~2391/2602 (and
consistent with the LDB age) or of a similar age. Note that any
uncertainty in the reddening tends to shift points parallel to the
cluster sequence, so is relatively unimportant compared with the
distance uncertainty. There is some indication from the upper part of
the diagram, where the 30 and 100\,Myr isochrones converge,
that the assumed distance of 370\,pc for IC~4665 may be too
small. Better agreement with the shape of the isochrones would be
achieved with a larger distance (say 400\,pc), which would move the
IC~4665 towards a younger isochronal age, in accord with its LDB
age and younger than IC~2391/2602.
 
The ages determined from the nuclear turn-off in the
Hertzsprung-Russell diagram do not contradict the age order
of these clusters determined from their LDBs.
IC~4665, IC~2391 and IC~2602 were assigned nuclear
turn-off ages of 36\,Myr by Mermilliod (1981) who also assigned the Pleiades
an older age of 78\,Myr.  These ages are dependent on
uncertain interior physics such as the amount of convective core
overshoot, but the age sequence should be secure.

Other indicators of youth such as the fraction of G-stars which are
rapid rotators and the colour/spectral type at which H$\alpha$ first
appears in emission are less precise (e.g. Jeffries, Totten \& James
2000). In both these respects, IC~4665 appears as young as
IC~2391 and IC~2602, although the statistics are poor.  In the range
$0.6<(V-I)_0<1.1$, where significant spin down might be expected
between ages of 30 and 50\,Myr, there are 16/32 stars in IC~2391 and
IC~2602 (of those plotted in Fig.~\ref{liviplot}) that have $v \sin i
>20$\,km\,s$^{-1}$, where $v\sin i$ measurements come from the same
papers as the Li data.  In the Pleiades this fraction has dropped to
only 6/56, where we have taken $v\sin i$ from Queloz
et al. (1998), Soderblom et al. (1993a) or Stauffer \& Hartmann (1987),
in that priority order.  The equivalent fraction among IC~4665
candidate members is 5/15, so is similar to IC~2391/2602, but
inconsistent with the Pleiades (excluded with 95 per cent confidence by
a two-tailed Fisher test).  In IC~2391/2602 H$\alpha$ emission first
occurs at $(V-I)_0\simeq 1$ and is ubiquitous for $(V-I)_0>1.3$ (see
Stauffer et al. 1997; Jeffries et al. 2000).  
In IC~4665 H$\alpha$ emission is first seen in
a star with $(V-I)_0=0.84\pm 0.06$ and 15/19 stars with $(V-I)_0>1.2\pm
0.06$ exhibit H$\alpha$ emission (see Fig.~\ref{vvi2}).  Given the
likelihood that a few contaminating dwarfs remain in the sample, the
four cool candidates exhibiting H$\alpha$ absorption are possibly
non-members as discussed in section~\ref{membership}.

We conclude that the isochronal age, the rotation of solar type stars
and the chromospheric activity are consistent with the LDB age of
IC~4665 and support its ranking within the age sequence of young
clusters.

\subsection{Strong Li depletion in the K- and M-stars of IC 4665?}

Given the general agreement between the sequencing of the isochronal,
nuclear turn-off and LDB ages, and the likelihood that there is little
theoretical uncertainty in the latter, we are left with the task of
explaining why, in IC~4665, the Li-depletion pattern seen among its
K/M-stars would indicate an age that is significantly older. Even
though only a small number of stars have been observed, one would
estimate that IC~4665 is at least as old as the Pleiades from the
evidence presented in Figs.~\ref{liviplot}b and~\ref{abunplot}b.

One ``uninteresting'' possibility that can be discounted is that we
have not observed any members of IC~4665 with $1<(V-I)_0<2$ at all, and
that the small Li\,{\sc i}~EWs in Fig.~\ref{liviplot} have been
measured in contaminating field stars that randomly have the cluster
RV. There are three strong arguments against this: (i) the predicted
number of field star contaminants ($16\pm 3$ minus the 12 definite
non-members, minus the 4 ``possible members'') remaining in the whole candidate
sample is less than the number of cluster candidates in this colour
range; (ii) six candidates in this colour range exhibit H$\alpha$
emission, which is a strong indicator of youth and rarely seen in field
stars with these colours; (iii) the presence of Li-rich mid/late
M-dwarfs beyond the LDB in IC~4665 (see M08) conclusively demonstrates
that the cluster mass function is not truncated below G-dwarfs.  The
limits of $1<(V-I)_0<2$ correspond approximately to
$1.0>M/M_{\odot}>0.7$ (for a 28\,Myr Baraffe et al. 2002 isochrone).
de Wit et al. (2006) used a photometric survey and statistical
subtraction of background contamination to estimate that the mass
function of IC~4665, within 0.5 degrees of the cluster centre, is
$dN/dM \simeq 80$ at these masses.  Given that our survey has covered
87 per cent of ``eligible'' cluster members, with a further $\sim 13$
per cent lost through our RV selection (because only 87 per cent will be
within 1.5 sigma of the mean),
we would expect to have seen $\sim 20$ cluster members with
$0.7<M/M_{\odot}<1.0$, which is comparable to (and certainly not lower)
than the 10 candidates we have actually found.

More interesting physical explanations could centre around the chemical
composition and rotation rates in IC~4665.  Theoretically,
PMS Li-depletion among stars with $M>0.5\,M_{\odot}$ should be
extremely sensitive to radiative opacity in the outer envelope and
hence stellar metallicity. A higher metallicity should lead to deeper
convection zones with hotter base temperatures and hence more extensive
Li burning (see Pinsonneault 1997). 
For instance Piau \& Turck-Chi\`eze (2002) show that a mere
6 per cent increase in average metallicity could increase the PMS
Li-depletion in solar mass stars by a factor of two. Likewise, the
models of Siess, Dufour \& Forestini (2000) predict that a
0.9\,$M_{\odot}$ star with a metallicity of $Z=0.02$ will have depleted
lithium by 1.2\,dex at an age of 30\,Myr, but if the metallicity is
increased to $Z=0.03$ the star {\it has a similar $T_{\rm eff}$} but has
depleted lithium by more than 5 dex!

At $\simeq 30$\,Myr a 0.9\,$M_{\odot}$ star has $(V-I)_0 \simeq 1.3$
and $T_{\rm eff}\simeq 4300$\,K according to the Baraffe et al. (2002)
models we are using. Figures~\ref{liviplot} and~\ref{abunplot} show
that in order to place the K-stars of IC~4665 above those of
IC~2391/2602 or the Pleiades would require about 1\,dex less
Li-depletion and about 2\,dex less Li-depletion in order to agree with the
predictions of the corresponding Baraffe et al. (2002) model isochrone.
Perhaps this could be explained if IC~4665 were modestly more
metal-rich than IC~2602/2391, the Pleiades and the metallicity adopted
in the Baraffe et al. models?  The available evidence does not wholly
support or exclude this.  Shen et al. (2005) spectroscopically
determined [Fe/H]$=-0.03\pm 0.04$ for F- and G-stars in IC~4665,
with the abundances of other elements also similar to solar values.
Platais et al. (2007) found [Fe/H]$=+0.06\pm0.06$ for IC~2391, whilst
the Pleiades has [Fe/H]$=-0.10 \pm 0.02$ and solar abundance ratios
(Wilden et al. 2002), both determined using similar techniques. 

A recent paper by D'Orazi \& Randich (2009) obtains even more precise
values of [Fe/H]$=-0.01 \pm 0.02$ and $0.00\pm 0.01$ for IC~2391 and
IC~2602 respectively, with the abundance ratios for other elements
close to solar values too. It does not seem likely therefore that
IC~4665 is more metal-rich than IC~2391 and IC~2602 unless there is
some large discrepancy in the abundance analysis methodology of Shen et
al. (2005) and D'Orazi \& Randich (2009). Both analyses used Kurucz
model atmospheres and derive abundances differentially with respect to
the Sun. However, Shen et al. (2005) fit their spectra with synthetic
models to simultaneously derive stellar parameters (including $T_{\rm
eff}$) and abundances, whilst D'Orazi \& Randich derive $T_{\rm eff}$
from photometric indices and then obtain abundances from a more
traditional equivalent width analysis.  A possible source of systematic
error is therefore the adopted temperature scale, but Shen et al. show
that their fitted $T_{\rm eff}$ values are in excellent agreement with
values determined from the relationship between $T_{\rm eff}$ and
$(B-V)_0$ proposed by Alonso, Arribas \& Mart\'inez-Roger (1996) using
$E(B-V)=0.18$ for IC~4665. This relationship is in turn very similar to
the $T_{\rm eff}/(B-V)_0$ relationship used by D'Orazi \& Randich
(2009) (within 30\,K at 5500\,K), so no inconsistency is indicated. It
would be a valuable exercise to analyse the IC~2391/2602 spectra using
the same approach adopted by Shen et al. (2005) to rule out any
remaining possibility of systematic error.

In summary, IC~4665 could be of order 0.1\,dex more metal-rich
than the Pleiades -- hence explaining the relative Li abundances of the
K- and M-stars of these clusters. However, on the basis of the
published evidence, IC~4665 seems unlikely to be more metal-rich than
the IC~2391 and IC~2602 clusters by more than a few hundredths of a
dex. If this were responsible for the excessive Li depletion of K- and
M-stars in IC~4665 with respect to the predictions of PMS models, then
the influence of metallicity must be very strong indeed. The lack of
significant differences between the Li abundance trends seen in a
number of Pleiades-age clusters argues against such a strong
metallicity dependence during the PMS Li-depletion phase 
(e.g. Soderblom et al. 1993b; Jeffries, James \& Thurston 1998; 
Jeffries \& James 1999; Barrado y Navascu\'es et al. 2001).

\begin{figure}
\centering
\includegraphics[width=80mm]{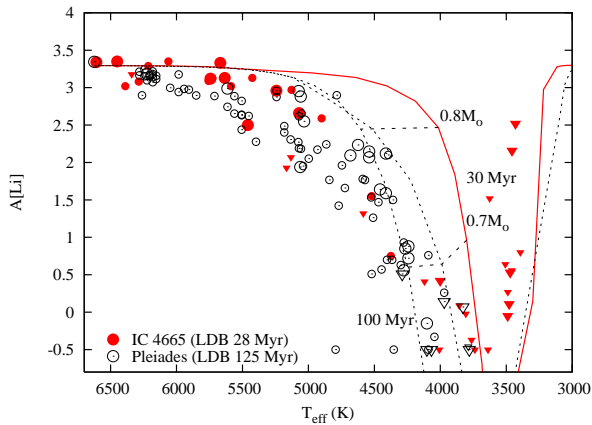}
\caption{A comparison of IC~4665 with the Pleiades in the
  $A$(Li)-$T_{\rm eff}$ plane, where fast rotators ($v \sin i >
  20$\,km\,s$^{-1}$) are given larger symbols. The plot also shows
  evolutionary tracks (from Baraffe et al. 2002) for stars of 0.7
  and 0.8\,$M_{\odot}$ to demonstrate that they have almost completed
  their Li-depletion by ages of 30\,Myr and their evolution in this
  diagram is governed mainly by increasing $T_{\rm eff}$ (see text).
}
\label{rotliabun}
\end{figure}

Another possibility is a dependence of Li-depletion on rotation rate.
Several authors have remarked that the wide spread in Li\,{\sc i} EWs
and consequently deduced Li abundances among the K-stars of young
clusters may be correlated with rotation rate. This finding is based
mainly on the Pleiades and clusters of similar age (e.g. Soderblom et
al. 1993b; Jeffries et al. 1998; Randich et al. 1998), where it is
found that the fastest rotators {\it tend} to have the strongest
Li\,{\sc i} features and the lower boundary of the $A$(Li) distribution
is defined {\it mainly} by the slowest rotators. This may indicate that
the fast rotators have preserved more of their initial Li, but could
also be caused by problems with the model atmospheres, associated with
rotationally-induced chromospheric activity or starspots (Stuik, Bruls \&
Rutten 1997; Ford, Jeffries \& Smalley 2002).

It is notable in this context that only 1 out of 10 IC~4665 candidates
with $1<(V-I)_0 < 2$ has $v \sin i > 20$\,km\,s$^{-1}$.  In contrast,
the equivalent fractions for the IC~2391/2602 and Pleiades samples in
Fig.~\ref{liviplot} are 10/25 and 18/44. According to a two-tailed
exact Fisher test we can marginally reject the null hypothesis that the
fraction of rapid rotators is similar in IC~4665 to IC~2391/2602 or the
Pleiades at confidence levels of 88 or 92 per cent
respectively. Balanced against this, one might expect
0.7-1.0\,$M_{\odot}$ stars to spin-up as they contract between 28\,Myr
and 125\,Myr, but even neglecting angular momentum loss this spin-up
should only be a modest factor of $\sim 1.2$ (Baraffe et al. 2002).

Figure~\ref{rotliabun} represents this information in graphical form,
where the fast rotators in the Pleiades and IC~4665 are identified by
larger symbols. The Pleiades fast rotators appear to have
depleted less Li than their slower rotating counterparts and these in
turn have similar levels of Li depletion to the few K- and early M-type
stars in IC~4665. Our view of the IC~4665 Li-depletion pattern might
change if some rapidly rotating K-star members were found with
higher levels of Li, but as discussed above, our data should include 75
per cent of cluster members in our survey area. If the RV
selection criteria were relaxed so that 87 per cent of cluster members
would have been included, there are still no Li-rich stars to be found. 
Even if a rapidly rotating, Li-rich member should emerge from the few
remaining unstudied candidates, we would still have to explain why
there are K- and M-type stars in IC~4665 that appear as Li-depleted
as the most depleted stars of similar types 
in IC~2391, IC~2602 and the Pleiades. 

Finally we comment on the status of the PMS models themselves.  There
may be a number of modifications or additions required in order to
match emerging measurements of masses and radii for eclipsing PMS
binaries (e.g. see Stassun et al. 2004). Much of the discussion centres
around the efficiency of convection and whether it is simply less
effective than usually assumed (both on the PMS and for main-sequence
binaries) or whether it may be suppressed by the presence of
dynamo-generated fields at the interface between radiative and
convective zones (e.g. Chabrier, Gallardo \& Baraffe 2007).

The lack of progression with age in Fig.~\ref{abunplot} between
IC~4665, IC~2391/2602 and the Pleiades (and other clusters at a similar
age) offers an important constraint, but also presents an interesting
dilemma when considered in this context. The problem is not how to
prevent further Li depletion between ages of 30 and 125\,Myr, but to
explain why stars of a given Li abundance do not get significantly
hotter over this period. This is demonstrated in Fig.~\ref{rotliabun},
where evolutionary tracks for stars of 0.7 and 0.8\,$M_{\odot}$ are
shown. In the Baraffe et al. (2002) models with a relatively low
convection efficiency (a mixing length of 1.0 pressure scale height),
Li depletion is almost complete by an age of 30\,Myr at these
masses. The change from the 30\,Myr through to the 100\,Myr isochrone is
driven by the development of the radiative core, the corresponding
change in the average polytropic index of the star and its traverse
across the Hertzsprung-Russell diagram to hotter temperatures along the
Henyey track (Henyey, Lelevier \& Lev\'ee 1955). To prevent, or
minimise this evolution in $T_{\rm eff}$ requires {\em more} efficient
convection (or increased radiative opacities).  This is clearly seen in
current models with higher convective efficiency or higher metallicity
(opacity), where there is much less difference between the 30 and
100\,Myr $A$(Li)-$T_{\rm eff}$ isochrones (e.g. see Fig.~9 in Jeffries
et al. 2003).  But paradoxically, this leads to much greater Li
depletion at a given age and $T_{\rm eff}$, so that such models cannot
match the observed data, unless there were some compensatory effect,
for example in the atmospheric modelling, which made the models much
cooler (by several hundred Kelvin).

\section{Summary}

In this paper we have used a radial velocity survey to select candidate
members of the young cluster IC~4665. The candidates have been filtered
according to a number of criteria and the final list (Table~2) contains
40 highly probable members and a few possible members, 
of which we estimate only a few could still be contaminating
foreground dwarfs. Table~2 should contain 75 per
cent of cluster members with $11.5<V<18$ over the central square degree
of the cluster.

The nuclear turn-off and isochronal ages of IC~4665 are consistent with
its lithium depletion boundary age of $28\pm5$\,Myr.  However, the
amount of lithum depletion seen among the K- and early-M stars of
IC~4665 is greater than expected when compared with similar stars in other
clusters of known age, and if judged in isolation, would
indicate an age as old as the Pleiades at 125\,Myr.  Despite
the small number of stars observed, this finding has important
implications for the use of Li-depletion as an age indicator. It is
precisely in the regime of small number statistics for loose kinematic
groups or even individual field stars that this technique has been most
widely used in the literature to estimate ages (e.g. Zuckerman \& Song
2004; Barrado y Navascu\'es 2006; Mentuch et al. 2008). Our results
suggest that Li-depletion among young late-type stars is a very crude
age indicator indeed, with at least one other parameter besides age
determining the rate and overall level of PMS Li depletion.

A possible explanation could be differences in metallicity and hence
radiative opacities between different clusters/stellar groups, leading
to variation in convection zone depth and PMS Li depletion. If so, then
even metallicity differences similar in size to their systematic
uncertainties could be very important. Only a very careful, homogeneous
study of Li-depletion in young clusters of similar age, but differing
metallicities could confirm and calibrate this effect.  Alternatively,
there are hints that the rotation rates of K- and M-stars in IC~4665
are slower than in comparison clusters and this may contribute to their
apparent Li depletion. Demonstrating a causal effect will be difficult
given the small number of K- and early M-type stars in IC~4665.

\section{Acknowledgements}
Based on observations made with the William Herschel Telescope operated
on the island of La Palma by the Isaac Newton Group in the Spanish
Observatorio del Roque de los Muchachos of the Instituto de
Astrofisica de Canarias. RJJ acknowledges receipt of a Science and
Technology Facilities Council postgraduate studentship. DJJ and PAC
acknowledge support from the National Science Foundation Career Grant
AST-0349075 (Principal Investigator: K.~G.~Stassun). Thanks are due to
an anonymous referee who helped us improve the manuscript.

\nocite{stauffer97}
\nocite{henyey55}
\nocite{chabrier07}
\nocite{stassun04}
\nocite{queloz98}
\nocite{stauffer87}
\nocite{fordpleiades02}
\nocite{stuik97}
\nocite{randich98}
\nocite{randich97}
\nocite{zuckerman04}
\nocite{barrado06}
\nocite{barrado01}
\nocite{jeffriesn251698}
\nocite{jeffriesblanco199}
\nocite{dorazi09}
\nocite{alonso96}
\nocite{manzi08}
\nocite{nidever02}
\nocite{mentuch08}
\nocite{jeffriescargese01}
\nocite{burke04}
\nocite{stauffer98b}
\nocite{barradoldb04}
\nocite{jeffries05}
\nocite{hogg55}
\nocite{crawford72}
\nocite{bessell98}
\nocite{landolt92}
\nocite{prosser93ic4665}
\nocite{prosser94ic4665}
\nocite{martin97}
\nocite{baraffe02}
\nocite{patten96}
\nocite{prosser96}
\nocite{randich01}
\nocite{jeffries03}
\nocite{jeffries09}
\nocite{sestito05}
\nocite{stauffer82}
\nocite{stauffer84}
\nocite{bessell87}
\nocite{mermilliod81}
\nocite{jeffries00}
\nocite{dewit06}
\nocite{shen05}
\nocite{wilden02}
\nocite{platais07}
\nocite{soderblom93pleiadesli}
\nocite{piau02}
\nocite{duncan83}
\nocite{stauffer95}
\nocite{song00}
\nocite{favata98}
\nocite{stetson87}
\nocite{smalley01}
\nocite{kurucz93}
\nocite{pavlenko96}
\nocite{jones96pleiades}
\nocite{zacharias04}
\nocite{garcia94}
\nocite{soderblom93rot}
\nocite{kenyon95}
\nocite{cutri03}
\nocite{carlsson94}
\nocite{siess00}
\nocite{vanleeuwen07}
\nocite{allain96}
\nocite{palla00}
\nocite{jeffries07b}
\nocite{horne86}
\nocite{pinsonneault97}
\nocite{cayrel88}
\nocite{pinsonneault90}
\nocite{anders89}

\label{lastpage}
\bibliographystyle{mn2e}  
\bibliography{iau_journals,master}

\end{document}